\documentclass[journal]{IEEEtran}
\usepackage{cite}

\ifCLASSINFOpdf
   \usepackage[pdftex]{graphicx}
\else
   \usepackage[dvips]{graphicx}
\fi

\usepackage{amsfonts}
\usepackage[cmex10]{amsmath}

\usepackage{multirow}

\usepackage[usenames]{color}

\newcommand{\conj}[1]{#1^{\ast}} 													
\newcommand{\trsp}[1]{#1^T} 															
\newcommand{\hadamard}{\circ}

\DeclareMathOperator*{\argmax}{arg\,max} 									
\newcommand{\DFT}{\mathbf{F}}
\newcommand{\IDFT}{\conj{\DFT}}
\newcommand{\set}[1]{\left\{#1\right\}}
\newcommand{\stTrMtx}{\mathbf{T}} 												
\newcommand{\exc}{\backslash} 														
\newcommand{\C}{\mathbb{C}} 															
\newcommand{\abs}[1]{|#1|} 															
\newcommand{\card}[1]{\left|#1\right|} 										
\newcommand{\stdBasis}{\mathbf{e}} 												

\newcommand{\pdf}[2][]{p_{#1}(#2)} 													
\newcommand{\Prb}[2][]{P_{#1}(#2)}													
\newcommand{\N}[1]{\mathfrak{N}(#1)} 													
\newcommand{\E}[2][]{\mathbb{E}_{#1}\{#2\}} 								
\newcommand{\varOp}[2][]{\mathbb{V}_{#1}\{#2\}}							

\newcommand{\msg}[2]{\mu_{#1 \rightarrow #2}} 
\newcommand{\neigh}[1]{\text{ne}\left(#1\right)}
\newcommand{\est}[1]{\hat{#1}} 
\newcommand{\map}[1]{\hat{#1}} 

\newcommand{\vc}[1]{\mathbf{#1}}
\newcommand{\grkvc}[1]{\boldsymbol{#1}} 							
\newcommand{\frqvc}[1]{\boldsymbol{#1}} 							
\newcommand{\chTime}{h}
\newcommand{\nTime}{n}
\newcommand{\nTones}{N}
\newcommand{\nDataTones}{\nTones_{\mathsf{d}}}
\newcommand{\nPilotTones}{\nTones_{\mathsf{p}}}
\newcommand{\nNullTones}{\nTones_{\mathsf{n}}}
\newcommand{\dataSet}{\mathcal{D}}
\newcommand{\pilotSet}{\mathcal{P}}
\newcommand{\nullSet}{\mathcal{N}}
\newcommand{\constl}{\mathbb{S}}

\newcommand{\sym}{\mathsf{S}}
\newcommand{\sigTime}{s}
\newcommand{\pilot}{\mathsf{p}}
\newcommand{\pilotPwr}{\rho_{\pilot}}
\newcommand{\rSig}{y}
\newcommand{\rSigFrq}{\mathsf{Y}}

\newcommand{\chFrq}{\mathsf{H}}
\newcommand{\nFrq}{\mathsf{N}}
\newcommand{\nMixComp}{K}
\newcommand{\mixPrb}{\pi}
\newcommand{\var}{\gamma}
\newcommand{\nState}{z}
\newcommand{\bckN}{g}
\newcommand{\bckPwr}{\var^{(0)}}
\newcommand{\impN}{i}
\newcommand{\impNFrq}{\mathsf{I}}
\newcommand{\bckNFrq}{\mathsf{G}}
\newcommand{\gamp}[1]{\text{GAMP}(#1)}
\newcommand{\JCISB}[1]{\text{JCISB}\left(#1\right) }

\newcommand{\symBlf}{\beta} 																												
\newcommand{\posSymBlf}{\lambda} 																										
\newcommand{\noiseParams}{\theta_{\vc{\impN}}}
\newcommand{\chParams}{\theta_{\vc{\chTime}}}
\newcommand{\uTones}{\mathcal{U}} 																									
\newcommand{\notuTones}{\overline{\mathcal{U}}} 																									
\newcommand{\chgamp}[1][\uTones]{\gamp{\frqvc{\rSigFrq}_{#1},\frqvc{\chFrq}_{#1},\sqrt{N}\DFT_{#1},\vc{\chTime}}} 	
\newcommand{\ngamp}[1][\uTones]{\gamp{\frqvc{\rSigFrq}_{#1},\frqvc{\impNFrq}_{#1},\DFT_{#1},\vc{\impN}}} 
\newcommand{\coh}{\mu} 																															
\newcommand{\nInfoBits}{M_{\mathsf{i}}}					 																		
\newcommand{\nCodedBits}{M_{\mathsf{c}}} 																						
\newcommand{\nOfdmSyms}{Q} 																													
\newcommand{\infoBit}{b} 																														
\newcommand{\codedBit}{c} 																													
\newcommand{\symMap}{\mathcal{M}} 																									
\newcommand{\chPrf}{\nu}																														
\newcommand{\SER}{\textsf{SER} }
\newcommand{\BER}{\textsf{BER} }
\newcommand{\SNR}{\textsf{SNR} }
\newcommand{\MSE}{\textsf{MSE} }
\newcommand{\NMSE}{\textsf{NMSE} }

\newcommand{\putTable}[3]{\begin{table}[t]
                            \caption{#2}
                            \label{tab:#1}
                            \centering
                            #3
                          \end{table} }

\begin{document}
%
\title{A Factor Graph Approach to\\ Joint OFDM Channel Estimation and Decoding\\ in
Impulsive Noise Environments}

\author{Marcel~Nassar,~\IEEEmembership{Student~Member,~IEEE,}
Philip~Schniter,~\IEEEmembership{Senior~Member,~IEEE,}
        and~Brian~L.~Evans,~\IEEEmembership{Fellow,~IEEE}
\thanks{M. Nassar and B.L. Evans are with Wireless Networking and Communications Group (WNCG) part of the Department of Electrical and Computer Engineering at The University of Texas at Austin, Austin, TX 78712 USA (email: mnassar@utexas.edu, bevans@ece.utexas.edu).}
\thanks{P. Schniter is with the Department of Electrical and Computer Engineering, The Ohio State University, Columbus, OH 43210 USA (email: schniter@ece.osu.edu).}
}

\markboth{Submitted to IEEE Transactions on Signal Processing}%
{Nassar \MakeLowercase{\textit{et al.}}: A Factor Graph Approach to OFDM Receivers in Impulsive Noise}

\maketitle

\begin{abstract}
We propose a novel receiver for orthogonal frequency division multiplexing (OFDM) transmissions in impulsive noise environments. 
Impulsive noise arises in many modern wireless and wireline communication systems, such as Wi-Fi and powerline communications, due to uncoordinated interference that is much stronger than thermal noise. 
We first show that the bit-error-rate optimal receiver jointly estimates the propagation channel coefficients, the noise impulses, the finite-alphabet symbols, and the unknown bits.
We then propose a near-optimal yet computationally tractable approach to this joint estimation problem using loopy belief propagation. 
In particular, we merge the recently proposed ``generalized approximate message passing'' (GAMP) algorithm with the forward-backward algorithm and soft-input soft-output decoding using a ``turbo'' approach.
Numerical results indicate that the proposed receiver drastically outperforms existing receivers under impulsive noise and comes within $1$ dB of the matched-filter bound.
Meanwhile, with $N$ tones, the proposed factor-graph-based receiver has only $O(N\log N)$ complexity, and it can be parallelized. 
\end{abstract}

\IEEEpeerreviewmaketitle

\section{Introduction}
\label{sec:Introduction}

\IEEEPARstart{T}{he} main impairments to a communication system, whether wireless or wireline, are due to multipath propagation through a physical medium and additive noise.  Multipath propagation is commonly modeled as a linear convolution that, in the slow-fading scenario, can be characterized by a channel impulse response $\set{\chTime_j}_{j=0}^{L-1}$ that is fixed over the duration of one codeword. 
In the well-known ``uncorrelated Rayleigh/Ricean-fading'' scenario, the (complex-baseband) channel ``taps'' $\chTime_j$ are modeled as independent circular Gaussian random variables, and in the well known ``additive white Gaussian noise'' (AWGN) scenario, the time-domain additive noise samples $\{\nTime_t\}_{\forall t}$ are modeled as independent circular Gaussian random variables \cite{Tse2005}.

\subsection{Motivation}
In this work, we focus on applications where the uncorrelated-Rayleigh/Ricean-fading assumption holds but the AWGN assumption does not.
Our work is motivated by
extensive measurement campaigns of terrestrial wireless installations 
wherein the additive noise is \textit{impulsive}, with peak noise amplitudes reaching up to $40$~dB above the thermal background noise level \cite{Blackard1993,Lauber1999,Sanchez1999,Sanchez2004,Nassar2011a,Nassar2011c}.  
The noise affecting powerline communications (PLC) has also been shown to be highly impulsive, as well as bursty \cite{Zimmermann2002,Nassar2011,Nassar2012}.

We restrict our attention to systems employing (coded or uncoded) orthogonal frequency division multiplexing (OFDM) \cite{Tse2005}, as used in many modern cellular wireless standards (e.g., IEEE802.11n and LTE) and PLC standards (e.g., PRIME and IEEE1901).
OFDM is advantageous in that it facilitates data communication across convolutive multipath channels with high spectral efficiency and low complexity.

The impulsivity of noise has particular consequences for OFDM systems.
Recall that, in conventional OFDM receivers, the time-domain received signal is converted to the frequency domain through a discrete Fourier transform (DFT) \cite{Tse2005}, after which each subcarrier (or ``tone'') is demodulated independently. 
Such tone-by-tone demodulation is in fact optimal with AWGN and perfect channel estimates \cite{Tse2005}, and is highly desirable from a complexity standpoint, since it leaves the DFT as the primary source of receiver complexity, and thus requires only $O(N\log N)$ multiplies per symbol for $N$ tones.
When the time-noise is impulsive, however, the corresponding frequency-domain noise samples will be highly dependent, and tone-by-tone demodulation is no longer optimal.
We are thus strongly motivated to find near-optimal demodulation strategies that preserve the $O(N\log N)$ complexity of classical OFDM.
In this work, we propose one such solution that exploits recent breakthroughs in loopy belief propagation.

\subsection{Prior Work}
\label{sec:PriorWork}

\subsubsection{OFDM Reception in Impulsive Noise}
\label{sec:ReceiversInImpulsiveNoise}

One popular approach to OFDM reception in impulsive noise stems from the argument that the noiseless time-domain received OFDM samples can be modeled as i.i.d Gaussian (according to the central limit theorem with sufficiently many tones), in which case the noise impulses can be detected using a simple threshold test. 
This approach straightforwardly leads to a decoupled procedure for impulse mitigation and OFDM reception: the time-domain received signal is \emph{pre-processed} via clipping or blanking techniques  
\cite{Zhidkov2008,Tseng2012} or (nonlinear) MMSE estimation \cite{Haring2001},  
and the result passed to a conventional DFT receiver for decoding.
While agreeable from a complexity standpoint, these techniques perform relatively poorly, especially when the power of the implusive noise is comparable to the power of the OFDM signal, or when higher order modulations are used \cite{Haring2001}.
This loss of performance can be explained by the fact that OFDM signal structure is not exploited for noise mitigation.
In an attempt to improve performance, it has been suggested to iterate between such pre-processing and OFDM decoding, but the approaches suggested to date (e.g., \cite{Haring2003,Mengi2010,Nassar2011b,Yih2012}) have shown limited success, mainly because the adaptation of preprocessing with each iteration is challenging and often done in an ad-hoc manner. 

Another popular approach models the time-domain impulsive noise sequence as a sparse vector and then uses \emph{sparse-reconstruction} techniques to estimate this sequence from the observed OFDM null and pilot (i.e., known) tones. 
The recovered impulse vector is then subtracted from the time-domain received signal, and the result is passed to a conventional DFT receiver for decoding.
Algebraic techniques were proposed in \cite{Wolf1983,Abdelkefi2005,Abdelkefi2007}, and sparse reconstruction techniques based on compressive-sensing were proposed in \cite{Caire2008,Lampe2011}. 
With typical numbers of known tones, these techniques have been shown to work well for very sparse impulsive noise sequences (e.g., one impulse in a $256$-tone OFDM system with $30$ known tones) but not for practical sparsity rates \cite{Caire2008,Lin2012}.

A more robust approach was proposed in \cite{Lin2012}, which performs joint symbol detection and impulse-noise estimation using sparse Bayesian learning (SBL).
Still, because \cite{Lin2012} decouples channel estimation from impulse-noise estimation and symbol detection, and because it integrates coding in an ad-hoc manner, there is considerable room for improvement. In addition, it performs matrix inversion that is impractical for typical OFDM receivers with hundreds of tones.

\subsubsection{Factor Graph Receivers}
\label{sec:FactorGraphReceivers}

Factor-graph-based receivers \cite{Worthen2001} have been proposed as a computationally efficient means of tackling the difficult task of \emph{joint} channel, symbol, and bit (JCSB) estimation.
Here, messages (generally in the form of pdfs) are passed among the nodes of the factor graph according to belief propagation strategies like the sum-product algorithm (SPA) \cite{Bishop2006}. 
Due to the loopy nature of the OFDM factor graph, however, exact implementation of the sum product algorithm is infeasible, and so various approximations have been proposed \cite{Novak2009,Liu2009,Kirkelund2010,Schniter2011}. 
Notably, \cite{Schniter2011} merged the ``generalized approximate message passing'' (GAMP) algorithm \cite{Rangan2011b} with a soft-input soft-output decoder in a ``turbo'' configuration to accomplish near-optimal\footnote{The approach was shown to be near-optimal in the sense of achieving \cite{Schniter2012} the pre-log factor of the sparse channel's noncoherent capacity \cite{Kannu2011}.} joint structured-sparse-channel estimation and decoding of bit-interleaved coded-modulation (BICM)-OFDM with $O(N\log N)$ complexity.
To our knowledge, no factor-graph-based OFDM receivers have been proposed to tackle impulsive noise, however.

\subsection{Contribution}
In this paper, we propose a novel OFDM receiver that performs near-optimally in the presence of impulsive noise while maintaining the $O(N\log N)$ complexity order of the conventional $N$-tone OFDM receiver.
Our approach is based on computing \textit{joint soft} estimates of the channel taps, the impulse-noise samples, the finite-alphabet symbols, and the unknown bits. 
Moreover, \textit{all} observed tones (i.e., pilots, nulls, and data tones) are exploited in this joint estimation. 
To do this, we leverage recent work on ``generalized approximate message passing'' (GAMP) \cite{Rangan2011b}, its ``turbo'' extension to larger factor graphs \cite{Schniter2010}, and off-the-shelf soft-input/soft-output (SISO) decoding \cite{MacKay2003}. 
The receiver we propose can be categorized as an extension of the factor-graph-based receiver \cite{Schniter2011} that explicitly addresses the presence of impulsive noise.
The resulting receiver provides a flexible performance-versus-complexity tradeoff and can be parallelized, making it suitable for FPGA implementations.

\subsection{Organization and Notation}
In Section~\ref{sec:SystemModel}, we describe our OFDM, channel, and noise models, and provide an illustrative example of impulsive noise. 
Then, in Section~\ref{sec:MessagePassingReceivers}, we detail our proposed approach, which we henceforth refer to as joint channel, impulse, symbol, and bit (JCISB) estimation. 
In Section~\ref{sec:NumericalResults} we provide extensive numerical results, 
and in Section~\ref{sec:Conclusion} we conclude.

\textit{Notation:}~
Vectors and matrices are denoted by boldface lower-case ($\vc{x}$) and upper-case notation ($\vc{X}$), respectively. $\vc{X}_{\mathcal{R},\mathcal{C}}$ then represents the sub-matrix constructed from rows $\mathcal{R}$ and columns $\mathcal{C}$ of $\vc{X}$, where the simplified notation $\vc{X}_{\mathcal{R}}$ means $\vc{X}_{\mathcal{R},:}$ and ``$:$'' indicates all columns of $\vc{X}$.  The notations $\trsp{(\cdot)}$ and $\conj{(\cdot)}$ denote transpose and conjugate transpose, respectively. The probability density function (pdf) of a random variable (RV) $X$ is denoted by $\pdf[X]{x}$, with the subscript omitted when clear from the context. Similarly, for discrete RVs, the probability mass function (pmf) is denoted by $\Prb[X]{x}$.  For a circular Gaussian RV with mean $\mu$ and variance $\var$, we write the pdf as $\N{x;\mu,\var}$.  The expectation and variance of a RV are then given by $\E{\cdot}$ and $\varOp{\cdot}$, respectively. 
We use the \textit{sans-serif} font to indicate frequency domain variables like $\mathsf{X}$ and bold \textit{sans-serif} to indicate frequency domain vectors like $\frqvc{\mathsf{X}}$.

\section{System Model}
\label{sec:SystemModel}

\subsection{Coded OFDM Model}
\label{sec:CodedOFDMModel}

We consider an OFDM system with $\nTones$ tones partitioned into 
$\nPilotTones$ pilot tones (indexed by the set $\pilotSet$), 
 $\nNullTones$ null tones (indexed by the set $\nullSet$), 
 and $\nDataTones$ data tones (indexed by the set $\dataSet$), each 
modulated by a finite-alphabet symbol chosen from an $2^M$-ary constellation $\constl$. 
The coded bits (which determine the data symbols) are generated by encoding $\nInfoBits$ information bits using a rate-$R$ coder, interleaving them, and allocating the resulting $\nCodedBits=\nInfoBits/R$ bits among an integer number $\nOfdmSyms = \left\lceil \nCodedBits/\nDataTones M \right\rceil $ of OFDM symbols.

In the sequel, we use $\sym^{(i)} \in \constl$ with $i\in\set{1,\dots,2^M}$ to denote the $i$th element of $\constl$, and $\vc{c}^{(i)} = \trsp{[c_1^{(i)},\dots,c_M^{(i)}]}$ to denote the corresponding bits as defined by the symbol mapping. 
Likewise, we use $\sym_k[q]$ to denote the scalar symbol transmitted on the $k$th tone of the $q$th OFDM symbol. 
Based on the tone partition, we note that: $\sym_k[q]=\pilot$ for all $k\in \pilotSet$, where $\pilot\in\C$ is a known pilot symbol; $\sym_k[q]=0$ for all $k\in \nullSet$; and $\sym_k[q]=\sym^{(l)}$ for some $l$ such that  $\vc{c}_k[q]=\vc{c}^{(l)}$ for all $k\in\dataSet$, where $\vc{c}_k[q]=\trsp{[c_{k,1}[q], \dots, c_{k,M}[q]]}$ denotes the coded/interleaved bits corresponding to $\sym_k[q]$. 
On the frame level, we use $\vc{c}[q]$ to denote the coded/interleaved bits allocated to the data tones of the $q$th OFDM symbol, and $\vc{c}=[\vc{c}[1],\dots,\vc{c}[Q]]$ to denote the entire codeword obtained from the information bits $\vc{b}=\trsp{[b_1,\dots,b_{\nInfoBits}]}$ by coding/interleaving. 
Similarly, we use $\frqvc{\sym}[q]=\trsp{[\sym_0[q], \dots, \sym_{N-1}[q]]}$ to denote the $q$th OFDM symbol's tone vector, including pilot, null, and data tones.

For OFDM modulation, an inverse of the unitary $\nTones$-point discrete Fourier transform (IDFT) matrix $\DFT$ is applied to the $q$th OFDM symbol's tone vector $\frqvc{\sym}[q]$, producing the time-domain sequence $\IDFT \frqvc{\sym}[q]=\frqvc{\sigTime}[q]=[\sigTime_0[q], \dots, \sigTime_{N-1}[q]]^T$, to which a cyclic prefix is prepended.
The resulting sequence propagates through an $L$-tap linear-time-invariant channel with impulse response $\vc{\chTime}[q]=\trsp{[\chTime_0[q], \dots, \chTime_{L-1}[q]]}$ before being corrupted by both AWGN and impulsive noise. 
Assuming a cyclic prefix of length $L\!-\!1$, inter-symbol interference is avoided by simply discarding the cyclic prefix at the receiver, after which the remaining $N$ samples are
\begin{equation}\label{eq:timeDomainRcvSig}
\vc{\rSig}[q] = \mathbf{H}[q]\vc{\sigTime}[q] + \vc{\nTime}[q] = \mathbf{H}[q]\IDFT \frqvc{\sym}[q] + \vc{\nTime}[q] 
\end{equation} 
where $\vc{\nTime}[q]$ is the time-domain noise vector 
and $\mathbf{H}[q]$ is the circulant matrix formed by $\vc{\chTime}[q]$ \cite{Tse2005}. 
Applying a DFT, the resulting frequency-domain received vector becomes
\begin{equation}\label{eq:frqDomainSystemEq}
\frqvc{\rSigFrq}[q]  
= \DFT\mathbf{H}[q]\IDFT \frqvc{\sym}[q] + \DFT \vc{\nTime}[q] =  \frqvc{\chFrq}[q] \hadamard \frqvc{\sym}[q] + \frqvc{\nFrq}[q]
\end{equation}
where $\frqvc{\chFrq}[q]= \sqrt{N}\DFT_{:,1:L}\vc{\chTime}[q]$ is the frequency-domain channel vector, $\frqvc{\nFrq}[q]=\DFT\vc{\nTime}[q]$ is the frequency-domain noise vector, and $\hadamard$ denotes the Hadamard (i.e., elementwise) product. 
The second equality in \eqref{eq:frqDomainSystemEq} follows from the fact that a circulant matrix is diagonalized by the Fourier basis.
In fact, \eqref{eq:frqDomainSystemEq} illustrates the principal advantage of OFDM: each transmitted tone $\sym_k[q]$ experiences a flat scalar subchannel, since
\begin{equation}\label{eq:subchannels}
\rSigFrq_k[q] = \chFrq_k[q] \sym_k[q] + \nFrq_k[q], \quad \forall k \in \{0,\dots,N-1 \}.
\end{equation} 

\subsection{Channel Modeling}
\label{sec:ChannelModeling}
We assume that the channel taps remain constant during the entire duration of one OFDM symbol, as required by \eqref{eq:frqDomainSystemEq}.
Since we make no assumptions on how the taps change across symbols, for simplicity we take $\vc{\chTime}[q]$ and $\vc{\chTime}[q']$ to be statistically independent for $q\neq q'$.
Furthermore, we use the Rayleigh-fading uncorrelated-scattering model 
\begin{equation}\label{eq:chPrior}
\{\chTime_j[q]\}_{q=-\infty}^{\infty} \sim \text{i.i.d~} \N{0,\chPrf_j}
\end{equation}
where $\trsp{[\chPrf_0,\dots ,\chPrf_{L-1}]}=\grkvc{\chPrf}$ is the power delay profile.
Extensions to sparse \cite{Schniter2012}, structured-sparse \cite{Schniter2011}, and time-varying sparse channels \cite{Schniter2011b} are straightforward, but not covered here.

\subsection{Impulsive Noise Models }
\label{sec:ImpulsiveNoiseModels}

\begin{figure}[!t]
\centering
\includegraphics[width=3in]{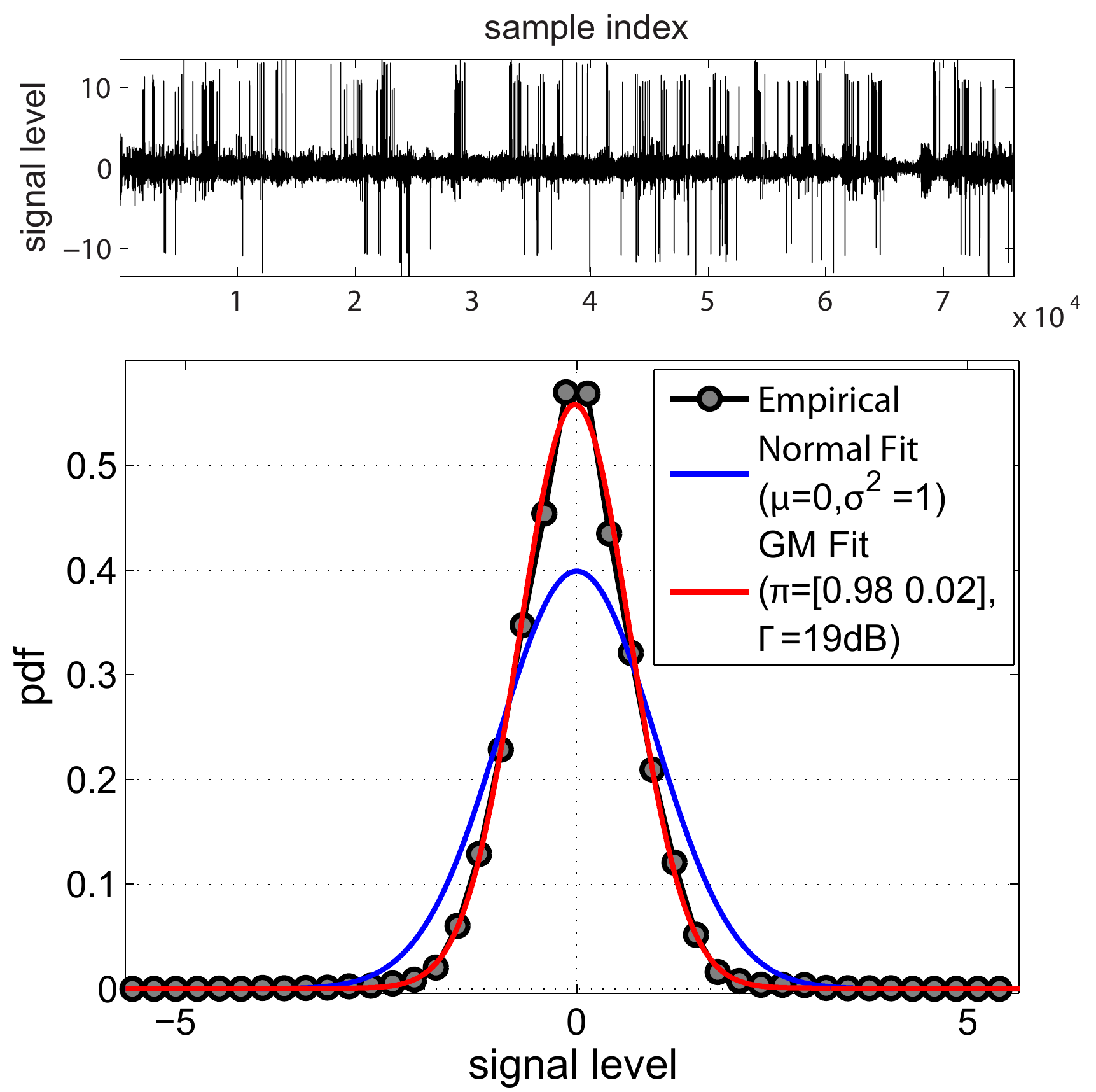}
\caption{Modeling a noise realization collected from a receiver embedded in a laptop: A $2$-component GM model provides a significantly better fit than a Gaussian model.}
\label{fig:GMTraceAndFit}
\end{figure}

In many wireless and power-line communication (PLC) systems, the additive noise is non-Gaussian (see the example in \figurename~\ref{fig:GMTraceAndFit}) and the result of random emission events from uncoordinated interferers (due to, e.g., aggressive spectrum reuse) or non-communicating electronic devices.  
In his pioneering work, Middleton modeled these random spatio-temporal emissions, or the ``noise field,'' using Poisson point processes (PPP), giving rise to the ``Middleton class-A'' and ``Middleton class-B'' noise models.
(For a recent review see \cite{Middleton1999}.) 
Recently, his approach has been extended to modeling fields of interferers in wireless and PLC networks using spatial and temporal PPPs, and 
the resulting interference was shown to follow either the Symmetric alpha stable, the Middleton class-A (MCA), or the more general Gaussian mixture (GM) distribution, depending on the network architecture \cite{Ilow1998,Gulati2010,Nassar2011,Nassar2012}. 
\figurename~\ref{fig:GMTraceAndFit} illustrates that a GM model provides a significantly better fit to a noise realization collected from a receiver embedded in a laptop than a Gaussian model does.

Since our factor-graph-based receiver is inherently Bayesian, these statistical models provide natural priors on the noise.
Thus, we model the additive noise using a GM model, noting that---given the pdf parameters---there is no distinction between the MCA and GM models.
In particular, we decompose\footnote{Our approach is equivalent to modeling the total noise $\nTime_t$ by a GM pdf $\pdf{\nTime_t} = \sum_{k=0}^{\nMixComp-1} \mixPrb^{(k)} \N{\nTime_t;0,\nu^{(k)}}$ with $\nu^{(0)}=\bckPwr$ and $\nu^{(k)}=\var^{(0)}+\var^{(k)}$ for $k>1$.} 
a given time-domain noise sample $\nTime_t = \bckN_t + \impN_t$ into a Gaussian background component $\bckN_t\sim \N{0,\bckPwr}$ and a sparse impulsive component $\impN_t$ with Bernoulli-GM pdf  
\begin{equation}\label{eq:impPdf}
\pdf{\impN_t} = \mixPrb^{(0)} \delta(\impN_t) + \sum_{k=1}^{\nMixComp-1} \mixPrb^{(k)} \N{\impN_t;0,\var^{(k)}} 
\end{equation}
where $\delta(\cdot)$ denotes the Dirac delta and $\sum_{k=0}^{K-1} \pi^{(k)}=1$.
Equivalently, we can model the (hidden) mixture state $\nState_t\in\{0,\dots,K\!-\!1\}$ of the impulsive component $\impN_t$ as a random variable, giving rise to the hierarchical model (with $\var^{(0)}=0$)
\begin{subequations}
\label{eq:impPdf2}
\begin{align}
p(\impN_t |\nState_t=k) &= \N{\impN_t;0,\var^{(k)}} \\
P(\nState_t=k) &= \mixPrb^{(k)} .
\end{align}
\end{subequations}

In many applications, such as PLC, the noise is not only impulsive but also \textit{bursty} and thus the noise samples are no longer statistically independent.
Such burstiness can be captured via a Bernoulli-Gaussian hidden Markov model (BGHMM) on the impulse-noise $\{\impN_t\}$ \cite{Zimmermann2002,Fruhwirth-Schnatter2006} or equivalently a Markov model on the GM state $\nState_t$ in \eqref{eq:impPdf2}.
For this, we model the sequence $\{\nState_t\}$ as a homogeneous (stationary) $K$-ary Markov chain with a state transition matrix $\stTrMtx$ such that
\begin{equation}
[\stTrMtx]_{i,j} = \Prb{\nState_{t}=j|\nState_{t-1}=i} \quad \forall i,j \in \set{0,\dots,\nMixComp-1} .
\end{equation}
In this case, the marginal pmf $\grkvc{\mixPrb}=[\mixPrb_0,\dots,\mixPrb_{K-1}]$ of steady-state $z_t$ obeys $\grkvc{\mixPrb}=\grkvc{\mixPrb}\stTrMtx$, 
and the mean duration of the event $\nState=k$ is $1/(1-\stTrMtx_{k,k})$ \cite{Fruhwirth-Schnatter2006}. 

\begin{figure}[!t]
\centering
\includegraphics[width=3.1in]{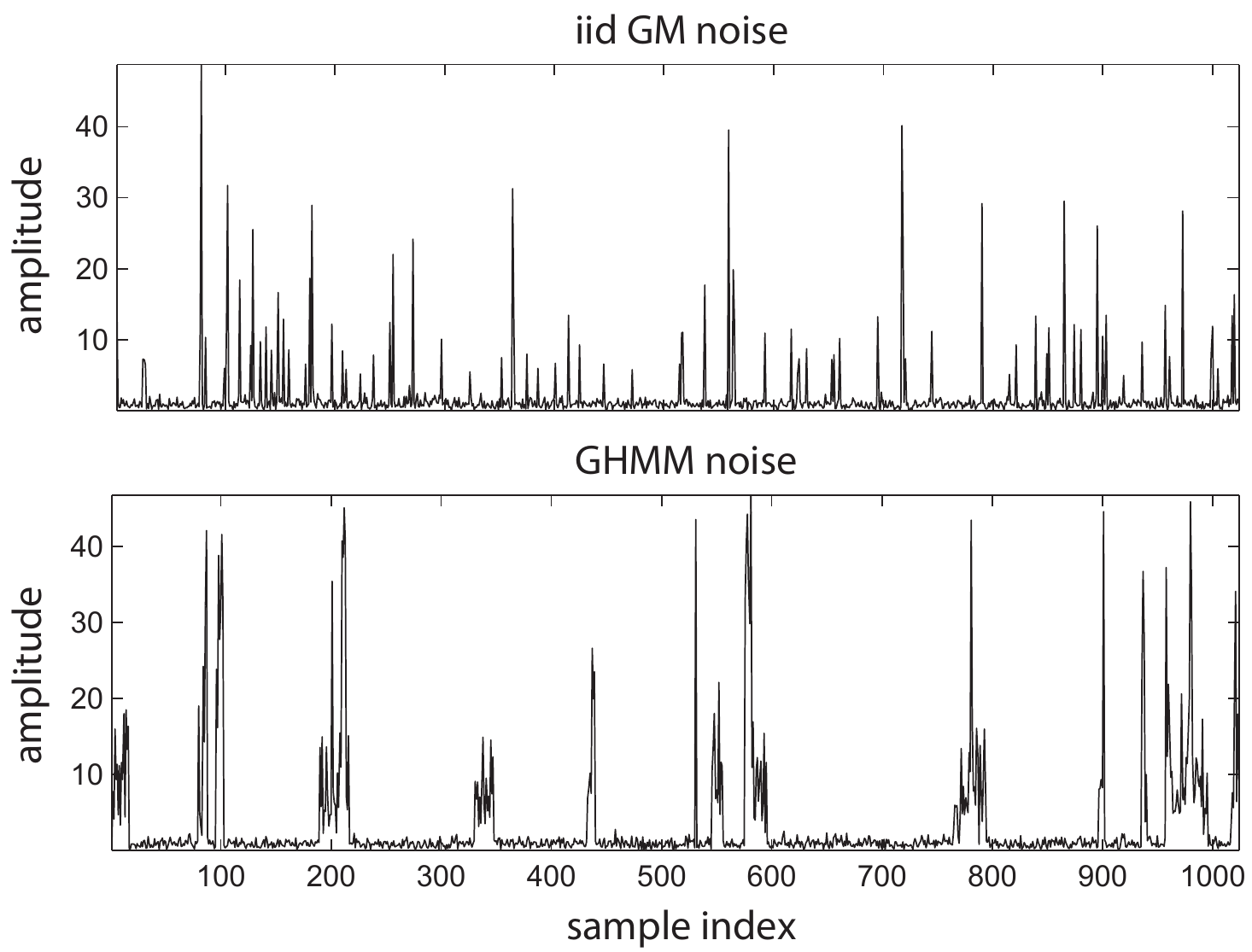}
\caption{Two realizations of noise $\{\nTime_t\}$ with identical marginals but different temporal statistics: the emission states $\{\nState_t\}$ in the top were generated i.i.d and those in the bottom were generated Markov via \eqref{eq:exampleTrsMatrix} to model burstiness.}
\label{fig:iidGmVSGHMMNoiseTrace}
\end{figure}

As an illustrative example, \figurename~\ref{fig:iidGmVSGHMMNoiseTrace} plots two realizations of the total noise $\{\nTime_t\}$ with impulsive component $\{\impN_t\}$ generated by the hierarchical Bernoulli-GM model \eqref{eq:impPdf2}.
Both realizations have identical marginal statistics: their impulsive components have two non-trivial emission states with powers $20$dB and $30$dB above the background noise power that occur $7\%$ and $3\%$ of the time, respectively.
However, in one case the emission state $\{\nState_t\}$ was generated i.i.d whereas in the other case it is generated Markov with state-transition matrix
\begin{equation}\label{eq:exampleTrsMatrix}
\stTrMtx=
\begin{bmatrix}
0.989 & 0.006 & 0.005 \\
0.064 & 0.857 & 0.079 \\
0.183 & 0.150 & 0.667
\end{bmatrix} .
\end{equation}
The GHMM realization clearly exhibits bursty behavior.

In practice, assuming the noise statistics are slowly varying, the noise parameters $\{\var_k\}_{k=0}^{K-1}$ and $\stTrMtx$ can be estimated using the expectation-maximization (EM) algorithm \cite{Bishop2006} during quiet intervals when there is no signal transmission. 

\section{Message-Passing Receiver Design}
\label{sec:MessagePassingReceivers}

In this section, we design computationally efficient message-passing receivers that perform near-optimal bit decoding which, as we shall see, involves jointly estimating the coded bits, finite-alphabet symbols, channel taps, and impulsive noise samples.
In doing so, our receivers exploit knowledge of the statistical channel and noise models discussed above and the OFDM signal structure (i.e., the pilot and null tones, the finite-alphabet symbol constellation, and the codebook).

\subsection{MAP Decoding of OFDM in Impulsive Noise}
\label{sec:OFDMDetectionInImpulsiveNoiseChannels}

Maximum a posteriori (MAP) decoding, i.e.,
\begin{equation}\label{eq:mapRule}
\map{\infoBit}_m = \argmax_{\infoBit_m \in \set{0,1}} \Prb{\infoBit_m|\frqvc{\rSigFrq}} \quad \forall m \in \set{1,\dots,\nInfoBits} 
\end{equation}
is well known to be optimal in the sense of minimizing the bit-error rate.
Here, $\frqvc{\rSigFrq}=[\frqvc{\rSigFrq}[1],\dots,\frqvc{\rSigFrq}[Q]]$ collects the received OFDM symbols of the corresponding frame.
Using the law of total probability, we can write the posterior information-bit probability from \eqref{eq:mapRule} as
\begin{align}
\lefteqn{ \Prb{\infoBit_m|\frqvc{\rSigFrq}} 
= \sum_{\vc{\infoBit_{\exc m}}} \Prb{\vc{\infoBit}|\frqvc{\rSigFrq}} \propto \sum_{\vc{\infoBit_{\exc m}}} \pdf{\frqvc{\rSigFrq}|\vc{\infoBit}} \Prb{\vc{\infoBit}} }\\
&\propto \sum_{\frqvc{\sym},\vc{\codedBit},\vc{\infoBit_{\exc m}}} \prod_{q=1}^Q \int_{\vc{\impN}[q],\vc{\chTime}[q]}
\pdf{\frqvc{\rSigFrq}[q]|\vc{\chTime}[q],\vc{\impN}[q],\frqvc{\sym}[q]} 
\nonumber\\&\quad\times
\pdf{\vc{\chTime}[q];\chParams}\pdf{\vc{\impN}[q];\noiseParams}\Prb{\frqvc{\sym}|\vc{\codedBit}} \Prb{\vc{\codedBit}|\vc{\infoBit}}
\\
&= \sum_{\frqvc{\sym},\vc{\codedBit},\vc{\nState},\vc{\infoBit_{\exc m}}} \prod_{q} \int_{\vc{\impN}[q],\vc{\chTime}[q]}
\prod_{k=0}^{N-1} \pdf{\rSigFrq_k[q]|\sym_k[q],\vc{\chTime}[q],\vc{\impN}[q]} 
\nonumber \\ & \quad \times 
\Prb{\sym_k[q]|\vc{\codedBit}_k[q]} 
\!\! \prod_{j=L}^{L+N-1} \!\!\pdf{\impN_j[q]|\nState_j[q]}
\,\Prb{\nState_j[q]|\nState_{j-1}[q]} 
\nonumber \\ & \quad \times 
\prod_{l=0}^{L-1} \pdf{\chTime_l[q]}
\,\Prb{\vc{\codedBit}|\vc{\infoBit}} 
\label{eq:posterior}
\end{align}
where ``$\propto$'' denotes equality up to a constant,
$\vc{\infoBit_{\exc m}}=\trsp{[\infoBit_1,\dots,\infoBit_{m-1},\infoBit_{m+1},\dots,\infoBit_{\nInfoBits}]}$,
and the information bits are assumed to be independent with $\Prb{\infoBit_m}=1/2~\forall m$. 
Equation \eqref{eq:posterior} shows that optimal decoding of $b_m$ involves marginalizing over the finite-alphabet symbols $\frqvc{\sym}$, coded bits $\vc{\codedBit}$, noise states $\vc{\nState}$, impulse noise samples $\vc{\impN}$, channel taps $\vc{\chTime}$, and other info bits $\vc{\infoBit_{\exc m}}$.

\begin{figure}[!t]
\centering
\includegraphics[width=3.493in]{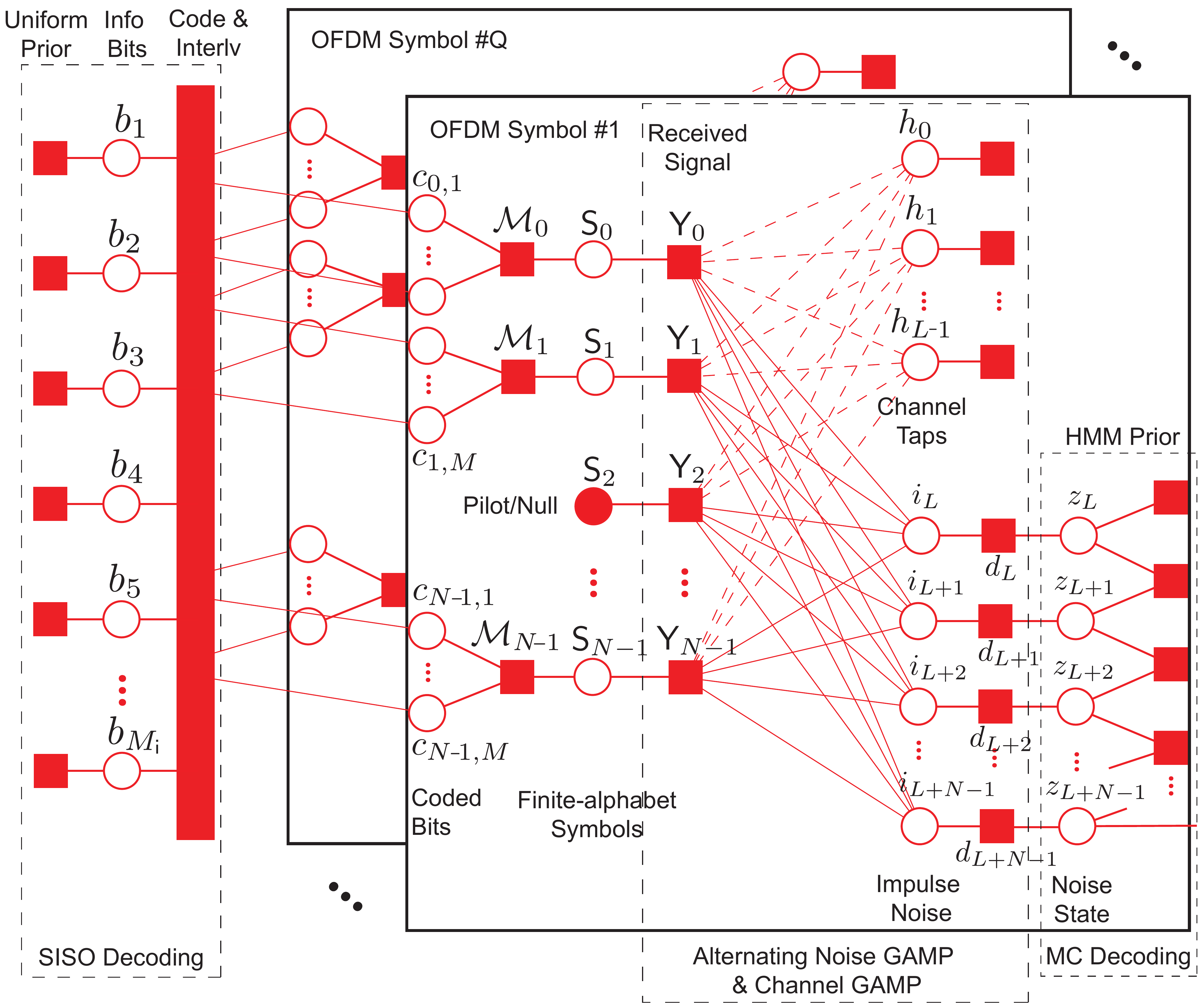}
\caption{Factor graph representation of a coded data frame spanning $Q$ OFDM symbols. 
Round open circles denote random variables, round solid circles denote deterministic variables (e.g., known pilots or nulls), and solid squares denote pdf factors.
The large rectangle on the left represents the coding-and-interleaving subgraph, whose details are immaterial. 
The time-domain impulse-noise quantities $i_t$ and $z_t$ start at time $t=L$ due to the use of an $L\!-\!1$-length cyclic prefix.}
\label{fig:factorSystemModel}
\end{figure}

The probalistic structure exposed by the factorization \eqref{eq:posterior} is illustrated by the \textit{factor graph} in \figurename~\ref{fig:factorSystemModel}. 
There and in the sequel, for brevity, we drop the ``$q$'' (i.e., OFDM symbol) index when doing so does not cause confusion.

Clearly, direct evaluation of $\Prb{\infoBit_m|\frqvc{\rSigFrq}}$ from \eqref{eq:posterior} is computationally intractable due to the high-dimensional integrals involved.
Belief propagation (BP), and in particular the sum-product algorithm (SPA) \cite{Bishop2006} described below, offers a practical alternative to direct computation of marginal posteriors.
In fact, when the factor graph has no loops, the SPA performs exact inference after only two rounds of message passing (i.e., forward and backward). 
On the other hand, when the factor graph is loopy, the computation of exact marginal posteriors is generally NP hard \cite{Cooper1990} and thus the posteriors computed by BP are generally inexact. 
Nevertheless, loopy BP has been successfully applied to many important problems, such as multi-user detection \cite{Boutros2002,Guo2007}, turbo decoding \cite{McEliece1998}, LDPC decoding \cite{MacKay2003}, compressed sensing \cite{Donoho2009,Rangan2011b}, and others.

In fact, for certain large densely loopy graphs that arise in the context of compressed sensing, SPA approximations such as the AMP \cite{Donoho2009} and GAMP \cite{Rangan2011b} algorithms are known to obey a state evolution whose fixed points, when unique, yield exact posteriors \cite{Bayati2011,Javanmard2012}.
Looking at the factor graph in \figurename~\ref{fig:factorSystemModel}, we see densely loopy sub-graphs between the factors $\{\rSigFrq_k\}$ and the time-domain noise samples $\{\impN_t\}$ and channel taps $\{\chTime_l\}$, which are due to the linear mixing of the Fourier matrix $\DFT$.  
It is these types of densely loopy graphs for which AMP and GAMP are designed.\footnote{Although rigorous GAMP guarantees have been established only for generalized linear inference problems with i.i.d sub-Gaussian transform matrices \cite{Javanmard2012}, equally good performance has been empirically observed over much wider classes of matrices \cite{Monajemi2013}.}
In the sequel, we will describe exactly how we combine the sum-product and GAMP algorithms for approximate computation of the bit posteriors in \eqref{eq:posterior}.
First, however, we review the SPA.

\subsection{Belief Propagation using Sum-Product Algorithm}
\label{sec:BeliefPropagationUsingSumProductAlgorithm}
Belief propagation (BP) transforms a high-dimensional marginalization problem (like \eqref{eq:posterior}) into a series of local low-dimensional marginalization problems by passing beliefs, or messages, which usually take the form of (possibly un-normalized) pdfs or pmfs, along the edges of a factor graph.
The sum-product algorithm (SPA) \cite{Bishop2006} is probably the best known approach to BP, and it operates according to the following rules:

\subsubsection{Messages from Factor Nodes to Variables}
\label{sec:MessagesFromFactorNodesToVariables}
Suppose the pdf factor $f_s(x_1,\dots,x_L)$ depends on the variables $\mathcal{X}_s=\{x_l\}_{l=1}^L$.
Then the message passed from factor node $f_s$ to variable node $x_m\in\mathcal{X}_s$ is 
\begin{equation*}
\msg{f_s}{x_m}(x_m) = \int_{\{x_l\}_{l\neq m}} \!\!\! f_s(x_1,\dots,x_L) \prod_{l\neq m} \msg{x_l}{f_s}(x_l)
\end{equation*}
representing the belief that node $f_s$ has about the variable $x_m$. 

\subsubsection{Messages from Variables to Factor Nodes}
\label{sec:MessagesFromVariablesToFactorNodes}
Suppose the factors $\mathcal{F}_m=\{f_1,\dots,f_M\}$ all involve the variable $x_m$.
Then the message passed from variable node $x_m$ to factor node $f_s\in\mathcal{F}_m$ is 
\begin{equation*}
\msg{x_m}{f_s}(x_m) = \prod_{k=1}^M \msg{f_k}{x_m}(x_m)
\end{equation*}
and represents the belief about the variable $x_m$ that node $x_m$ passes to node $f_s$.

\subsubsection{Marginal Beliefs}
\label{sec:MarginalApproximation}
SPA's approximation to the marginal posterior pdf on the variable $x_m$ is
\begin{equation*}
\pdf{x_m} = C \! \! \! \prod_{f_s \in \neigh{x_m}} \! \! \! \msg{f_s}{x_m} \left(x_m\right)
\end{equation*}
where $C$ is a normalization constant.

\subsection{Joint Channel/Impulse-Noise Estimation and Decoding}
\label{sec:JointChannelAndNoiseEstimationAndDecoding}
We now propose a strategy to approximate the bit posteriors in \eqref{eq:posterior} by iterating (an approximation of) the SPA on the loopy factor graph in \figurename~\ref{fig:factorSystemModel}.
To distinguish our approach from others in the literature, we will refer to it as ``joint channel, impulse, symbol, and bit estimation'' (JCISB).

Given the loopy nature of the factor graph, there exists considerable freedom in the message-passing schedule. 
In JCISB, we choose to repeatedly pass messages from right to left, and then left to right, as follows.
\begin{enumerate}
\item
Beliefs about coded bits $\{\codedBit_{k,m}\}$ flow rightward through the symbol-mapping nodes $\{\symMap_k\}$, the finite-alphabet symbol nodes $\{\sym_k\}$, and into the factor nodes $\{\rSigFrq_k\}$. 
\item
GAMP-based messages are then passed repeatedly between the $\{\rSigFrq_k\}$ and $\{\chTime_l\}$ nodes until convergence.
\item
GAMP-based messages are passed repeatedly between the $\{\rSigFrq_k\}$ and $\{\impN_t\}$ nodes until convergence, and then through the $\{\nState_t\}$ nodes using the forward-backward algorithm, alternating these two steps until convergence.
\item
Finally, the messages are propagated from $\{\rSigFrq_k\}$ leftward through the symbol nodes $\{\sym_k\}$, the symbol-mapping nodes $\{\symMap_k\}$, the coded-bit nodes $\{\codedBit_{k,m}\}$, and the coding-interleaving block---the last step via an off-the-shelf soft-input/soft-output (SISO) decoder.
\end{enumerate}
In the sequel, we will refer to steps 1)--4) as a ``turbo'' iteration, and to 
the iterations within step 3) as ``impulse iterations,''
We note that it is also possible to execute a parallel schedule if the hardware platform supports it.
The details of these four message passing steps are given below.

\subsubsection{Bits to Symbols}
Beliefs about the coded bits $\{\codedBit_{k,m}\}_{m=1}^M$ (for each data tone $k\in\dataSet$) are first passed through the symbol mapping factor node $\symMap_k$.
The SPA dictates that 
\begin{align}\label{eq:msgFromSymMapToSym}
\msg{\symMap_k}{\sym_k}(\sym^{(i)}) 
&= \sum_{\vc{\codedBit}_k\in\set{0,1}^M} \!\!\!\!\!\Prb{\sym^{(i)}|\vc{\codedBit}_k}
\prod_{m=1}^M\msg{\codedBit_{k,m}}{\symMap_k}(\codedBit_{k,m}) \nonumber \\
&= \prod_{m=1}^M \msg{\codedBit_{k,m}}{\symMap_k}(\codedBit_m^{(i)})
\end{align}
where \eqref{eq:msgFromSymMapToSym} follows from the deterministic symbol mapping $\Prb{\sym^{(i)}|\vc{\codedBit}^{(j)}} = \delta_{i-j}$. 
The resulting message is then copied forward through the $\sym_k$ node, i.e., 
$\msg{\sym_k}{\rSigFrq_k}=\msg{\symMap_k}{\sym_k}$, 
also according to the SPA.
Note that, at the start of the first turbo iteration, we have no knowledge of the bits and thus we take $\msg{\codedBit_{k,m}}{\symMap_k}(c)$ to be uniform across $c\in\{0,1\}$ for all $m,k$.

\subsubsection{GAMP for Channel Estimation}
The next step in our message-passing schedule is to pass messages between the factor nodes $\{\rSigFrq_k\}$ and the time-domain channel nodes $\{\chTime_l\}$.
According to the SPA, the message passed from $\rSigFrq_k$ to $\chTime_l$ is 
\begin{align}\label{eq:msgRSigFrq2ch}
\msg{\rSigFrq_k}{\chTime_l}(\chTime_l) &= \sum_{\sym_k} \int_{\vc{i},\vc{h}_{\exc l}}\pdf{\rSigFrq_k|\sym_k,\vc{\chTime},\vc{\impN}}\msg{\sym_k}{\rSigFrq_k}(\sym_k) \nonumber \\
 & \quad \times \prod_{z\neq l}\msg{\chTime_z}{\rSigFrq_k}(\chTime_z) \prod_{j}\msg{\impN_j}{\rSigFrq_k}(\impN_j).
\end{align}
Exact evaluation of \eqref{eq:msgRSigFrq2ch} involves an integration of the same high-dimensionality that made \eqref{eq:posterior} intractable, with exponential complexity in $N$. 
Thus, we instead approximate the message passing between the $\{\rSigFrq_k\}$ and $\{\chTime_l\}$ nodes using \emph{generalized approximate message passing} (GAMP) algorithm \cite{Rangan2011b} reviewed in Appendix~\ref{appndx:GAMPsummary} and summarized in Table~\ref{tab:gamp}.

\putTable{gamp}
{The $\gamp{\vc{y},\vc{z},\boldsymbol{\Phi},\vc{x}}$ Algorithm with 
coefficients-of-interest $\vc{x}$, 
linear transform $\boldsymbol{\Phi}$, 
transform outputs $\vc{z}=\boldsymbol {\Phi}\vc{x}$, 
and observed outputs $\vc{y}$}
{\footnotesize
\newcommand{\mc}[1]{\mathcal{#1}}
\renewcommand{\vec}[1]{\vc{#1}}
\renewcommand{\E}[1]{\mathbb{E}\{#1\}}
\renewcommand{\var}[1]{\mathbb{V}\{#1\}}
\renewcommand{\mu}{\gamma}
\setlength{\arraycolsep}{0.5mm}
\vspace{-4mm}
\begin{equation*}
\begin{array}{|lrcl@{}r|}\hline
  \multicolumn{4}{|l}{\textsf{inputs:~~}
        \{p(x_j)\}_{\forall j}, 
        \{p(y_i|z_i)\}_{\forall i}, 
        \vec{y},\vec{\Phi}, T_{\max}
        }&\\[1mm]
  \multicolumn{2}{|l}{\textsf{definitions:}}&&&\\[-1mm]
  &p(z_i|\vec{y};\hat{p},\mu^p)
   &=& \frac{p(y_i|z_i) \,\mc{N}(z_i;\hat{p},\mu^p)}
        {\int_{z_i'} p(y_i|z_i') \,\mc{N}(z_i';\hat{p},\mu^p)} &\text{(D1)}\\
  &p(x_j|\vec{y};\hat{r},\mu^r)
   &=& \frac{p(x_j) \,\mc{N}(x;\hat{r},\mu^r)}
        {\int_{x_j'}p(x_j') \,\mc{N}(x_j';\hat{r},\mu^r)}&\text{(D2)}\\
  \multicolumn{2}{|l}{\textsf{initialize:}}&&&\\
  &\forall j: 
   \hat{x}_j(1) &=& \int_{x_j} x_j\, p(x_j) & \text{(I1)}\\
  &\forall j:
   \mu^x_j(1) &=& \int_{x_j} |x_j-\hat{x}_j(1)|^2  p(x_j) & \text{(I2)}\\
  &\forall i: 
   \hat{s}_i(0) &=& 0 & \text{(I3)}\\
  \multicolumn{2}{|l}{\textsf{for $t=1:T_{\max}$,}}&&&\\
  &\forall i:
   \mu^p_i(t)
   &=& \textstyle \sum_{j=1}^{N} |\Phi_{ij}|^2 \mu^x_j(t) & \text{(R1)}\\
  &\forall i:
   \hat{p}_i(t)
   &=& \sum_{j=1}^{N} \Phi_{ij} \hat{x}_j(t) - \mu^p_i(t) \,\hat{s}_i(t-1)& \text{(R2)}\\
  &\forall i:
   \mu^z_i(t)
   &=& \var{z_i|\vec{y};\hat{p}_i(t),\mu^p_i(t)} & \text{(R3)}\\
  &\forall i:
   \hat{z}_i(t)
   &=& \E{z_i|\vec{y};\hat{p}_i(t),\mu^p_i(t)} & \text{(R4)}\\
  &\forall i:
   \mu^s_i(t)
   &=& \big(1-\mu^z_i(t)/\mu^p_i(t)\big)/\mu^p_i(t) & \text{(R5)}\\
  &\forall i:
   \hat{s}_i(t)
   &=& \big(\hat{z}_i(t)-\hat{p}_i(t)\big)/\mu^p_i(t) & \text{(R6)}\\
  &\forall j:
   \mu^r_j(t)
   &=& \textstyle \big(\sum_{i=1}^{M} |\Phi_{ij}|^2 \mu^s_i(t) 
        \big)^{-1} & \text{(R7)}\\
  &\forall j:
   \hat{r}_j(t)
   &=& \textstyle \hat{x}_j(t)+ \mu^r_j(t) \sum_{i=1}^{M} \Phi_{ij}^*
        \hat{s}_i(t)  & \text{(R8)}\\
  &\forall j:
   \mu^x_j(t\!+\!1)
   &=& \var{x_j|\vec{y};\hat{r}_j(t),\mu^r_j(t)} & \text{(R9)}\\
  &\forall j:
   \hat{x}_j(t\!+\!1)
   &=& \E{x_j|\vec{y};\hat{r}_j(t),\mu^r_j(t)} & \text{(R10)}\\
  \multicolumn{4}{|l}{\textsf{outputs:~~}
        \{\hat{z}_i(T),\mu^z_i(T)\}_{\forall i},
        \{\hat{r}_j(T),\mu^r_j(T),
          \hat{x}_j(T),\mu^x_j(T)\}_{\forall j} 
        }&\\[1mm]
  \hline
\end{array}
\end{equation*}
}

To do this, we temporarily treat the messages $\{\msg{\sym_k}{\rSigFrq_k}\}$ and $\{\msg{\impN_j}{\rSigFrq_k}\}$ as fixed, allowing us to employ ``$\chgamp[]$,'' 
using the notation established in the caption of Table~\ref{tab:gamp}.
Definition (D1) [later used in steps (R3)--(R4)] requires us to specify the likelihood $\pdf{\rSigFrq_k|\chFrq_k}$ relating the transform output $\chFrq_k$ to the corresponding observed output $\rSigFrq_k$. 
From \figurename~\ref{fig:factorSystemModel}, we see that there are two types of belief flowing into each $\rSigFrq_k$ node (apart from beliefs about $\{\chTime_l\}$) that determine this likelihood: 
beliefs about the symbols $\{\sym_k\}$, which we parameterize as $\grkvc{\symBlf}_k=[\symBlf_k^{(1)},\dots,\symBlf_k^{(\card{\constl})}]$ with $\symBlf_k^{(i)}=\msg{\sym_k}{\rSigFrq_k}(\sym^{(i)})$, and 
beliefs about the frequency-domain impulsive noise $\{\impNFrq_k\}$, which GAMP approximates as $\N{\impNFrq_k;\hat{\impNFrq}_k,\var^{\impNFrq}_k}$, where the values $\{\hat{\impNFrq}_k,\var^{\impNFrq}_k\}$ were computed by $\ngamp[]$ in the previous turbo iteration.\footnote{During the first turbo iteration, we use $\hat{\impNFrq}_k=0$ and $\var^{\impNFrq}_k=\var^{\impN}~\forall k$.} 
Here, $\frqvc{\impNFrq}=\DFT\vc{\impN}$ refers to the impulsive component of the frequency-domain noise $\frqvc{\nFrq}=\frqvc{\impNFrq}+\frqvc{\bckNFrq}$, with $\{\bckNFrq_k\}\sim\text{i.i.d}~\N{0,\gamma^{(0)}}$, so that [from \eqref{eq:frqDomainSystemEq}]
\begin{align}\label{eq:ofdmFrqDomain}
\frqvc{\rSigFrq}
&= \frqvc{\chFrq} \hadamard \frqvc{\sym} + \frqvc{\impNFrq} + \frqvc{\bckNFrq} .
\end{align}
From \eqref{eq:subchannels} and \eqref{eq:ofdmFrqDomain}, the $\chgamp[]$ likelihood is 
\begin{equation}
\label{eq:chgamplike}
\pdf{\rSigFrq_k|\chFrq_k} = 
\begin{cases}
\N{\rSigFrq_k;\pilot\chFrq_k+\hat{\impNFrq}_k,\var_k^{\impNFrq}\!+\!\bckPwr} & k\in\pilotSet \\
\sum\limits_{l=1}^{\card{\constl}} \symBlf_k^{(l)} \N{\rSigFrq_k;\sym^{(l)}\chFrq_k+\hat{\impNFrq}_k,\var_k^{\impNFrq}\!+\!\bckPwr} & k\in\dataSet \\
\end{cases}
\end{equation}
with the corresponding ``output MMSE estimation functions'' $\E{\chFrq_k|\frqvc{\rSigFrq};\est{p},\var^{p}}$ and $\varOp{\chFrq_k|\frqvc{\rSigFrq};\est{p},\var^{p}}$, as used in steps (R3)--(R4), specified in Table~\ref{tab:outputScalarEstimationFun}. 
(See Appendix~\ref{appdx:channelDerivation} for derivations). 

\begin{table*}[!t]

\renewcommand{\arraystretch}{1.3}
\caption{GAMP output MMSE estimation functions used in JCISB}
\label{tab:outputScalarEstimationFun}
\centering
\begin{tabular}{|c|c|c|}
\hline
\multirow{2}{*}{\textbf{Tone Type}} & \multicolumn{2}{ |c| }{$\chgamp[]$} \\
\cline{2-3}
 & $\E{\chFrq_k|\rSigFrq_k;\est{p},\var^{p}}$ & $\varOp{\chFrq_k|\rSigFrq_k;\est{p},\var^{p}}$ \\
\hline
\textbf{Pilot:} $k\in\pilotSet$ &  
$\est{p} + \var^{p}\conj{\pilot}(\rSigFrq_k - \hat{\impNFrq}_k-\pilot\est{p})/(\bckPwr + \var^{\impNFrq}_k+\pilotPwr\var^{p})$ 
& $\var^{p}(\bckPwr+\var^{\impNFrq}_k)/(\bckPwr + \var^{\impNFrq}_k+\pilotPwr\var^{p})$ \\
\hline
\textbf{Data:} $k\in\dataSet$ &  
$\begin{array}{c}
\est{p} + \sum_{l=1}^{\card{\constl}} \posSymBlf_k^{(l)} \frac{\var^{p}\sym^{*(l)}(\rSigFrq_k - \hat{\impNFrq}_k-\hat{p}\sym^{(l)})}{(\bckPwr+\var^{\impNFrq}_k+\abs{\sym^{(l)}}^2\var^{p})} \\
\text{ where } \posSymBlf_k^{(l)} =\pdf{\rSigFrq_k|\sym^{(l)}}\symBlf_k^{(l)}/\sum_j \pdf{\rSigFrq_k|\sym^{(j)}}\symBlf_k^{(j)} \text{ and }\\
\pdf{\rSigFrq_k|\sym^{(l)}} = \N{\rSigFrq_k;\hat{\impNFrq}_k+\hat{p}\sym^{(l)},\bckPwr+\var^{\impNFrq}_k+\abs{\sym^{(l)}}^2\var^{p}}
\end{array}$ & 
$\begin{array}{c}
\sum_{l=1}^{\card{\constl}}\posSymBlf_k^{(l)}\Big[\frac{\var^{p}(\bckPwr+\var^{\impNFrq}_k)}{\bckPwr+\var^{\impNFrq}_k+\abs{\sym^{(l)}}^2\var^{p}}  + \\
\Big|\hat{p}
+ \frac{\var^{p}\sym^{*(l)}(\rSigFrq_k-\hat{\impNFrq}_k -\hat{p}\sym^{(l)})}{(\bckPwr+\var^{\impNFrq}_k+\abs{\sym^{(l)}}^2\var^{p})} \Big|^2\Big] -\abs{\E{\chFrq_k|\rSigFrq_k;\est{p},\var^{p}}}^2
\end{array}$
 \\
\hline
\hline
 \multirow{2}{*}{\textbf{Tone Type}} & \multicolumn{2}{ |c| }{$\ngamp[]$} \\
\cline{2-3}
 & $\E{\impNFrq_k|\rSigFrq_k;\est{p},\var^{p}}$ & $\varOp{\impNFrq_k|\rSigFrq_k;\est{p},\var^{p}}$ \\
\hline
\textbf{Null:} $k\in\nullSet$ & $(\var^{p}\rSigFrq_k-\bckPwr\hat{p})/(\bckPwr+\var^{p})$ & $\var^{p}\bckPwr/(\bckPwr+\var^{p})$ \\
\hline
\textbf{Pilot:} $k\in\pilotSet$ &  $\est{p} + \var^{p}(\rSigFrq_k-\hat{p}-\hat{\chFrq}_k\pilot)/(\bckPwr+\var^{p}+\pilotPwr\var_k^{\chFrq})$ & $\var^{p}(\bckPwr+\pilotPwr\var_k^{\chFrq})/(\bckPwr+\var^{p}+\pilotPwr\var_k^{\chFrq})$ \\
\hline
\textbf{Data:} $k\in\dataSet$ & 
$\begin{array}{c}
\est{p}+\sum_{l=1}^{\card{\constl}} \posSymBlf_k^{(l)} \frac{\var^{p}(\rSigFrq_k - \hat{p}-\hat{\chFrq}_k\sym^{(l)})}{(\bckPwr+\var^{p}+\abs{\sym^{(l)}}^2\var_k^{\chFrq})} \\
\text{ where } \posSymBlf_k^{(l)} =\pdf{\rSigFrq_k|\sym^{(l)}}\symBlf_k^{(l)}/\sum_j \pdf{\rSigFrq_k|\sym^{(j)}}\symBlf_k^{(j)} \text{ and }\\
\pdf{\rSigFrq_k|\sym^{(l)}} = \N{\rSigFrq_k;\hat{p}+\hat{\chFrq}_k\sym^{(l)},\bckPwr+\var^{p}+\abs{\sym^{(l)}}^2\var^{\chFrq}_k}
\end{array}$ 
& 
$\begin{array}{c}
\sum_{l=1}^{\card{\constl}}\posSymBlf_k^{(l)}\Big[\frac{\var^{p}(\bckPwr+\abs{\sym^{(l)}}^2\var^{\chFrq}_k)}{(\bckPwr+\var^{p}+\abs{\sym^{(l)}}^2\var^{\chFrq}_k)} + \\  
\Big|\hat{p} +
\frac{\var^{p}(\rSigFrq_k-\hat{p} -\hat{\chFrq}_k\sym^{(l)})}{\bckPwr+\var^{p}+\abs{\sym^{(l)}}^2\var^{\chFrq}_k} \Big|^2\Big] 
-\abs{\E{\impNFrq_k|\rSigFrq_k;\est{p},\var^{p}}}^2
\end{array}$ \\ 
\hline
\end{tabular}
\end{table*}

$\chgamp[]$ also requires us to derive the ``input MMSE estimation functions'' $\E{\chTime_j|\frqvc{\rSigFrq},\est{r},\var^r}$ and $\varOp{\chTime_j|\frqvc{\rSigFrq},\est{r},\var^r}$ for GAMP steps (R9)--(R10).
Given the channel model specified in Section~\ref{sec:ChannelModeling} and definition (D2), it is straightforward to show \cite{Rangan2011b} that the input MMSE estimation functions are $\E{\chTime_j|\frqvc{\rSigFrq},\est{r},\var^r}=\chPrf_j \est{r}/(\chPrf_j+\var^r)$ and $\varOp{\chTime_j|\frqvc{\rSigFrq},\est{r},\var^r}=\chPrf_j \var^r/(\chPrf_j+\var^r)$.  

After $\chgamp[]$ is iterated to convergence, the outputs $\{\hat{\chFrq}_k\}$ and $\{\var^{\chFrq}_k\}$ of steps (R4)--(R3) are close approximations to the marginal posterior mean and variance, respectively, of $\{\chFrq_k\}$.
These outputs will be used in the next step of the message-passing schedule, as described below.
Similarly, the outputs $\{\hat{\chTime}_l\}$ and $\{\var^{\chTime}_l\}$ of steps (R10)--(R9) are close approximations to the marginal posterior mean and variance, respectively, of $\{\chTime_l\}$.

\subsubsection{Turbo-GAMP for Noise Estimation}

The next step in our schedule is to pass messages between the factor nodes $\{\rSigFrq_k\}$, the time-domain impulse-noise nodes $\{\impN_t\}$, and the noise-state nodes $\{\nState_t\}$.
According to the SPA, the message passed from $\rSigFrq_k$ to $\impN_t$ is 
\begin{align}\label{eq:msgRSigFrq2noise}
\msg{\rSigFrq_k}{\impN_t}(\impN_t) 
&= \sum_{\sym_k} \int_{\vc{i}_{\exc t},\vc{h}}\pdf{\rSigFrq_k|\sym_k,\vc{\chTime},\vc{\impN}}\msg{\sym_k}{\rSigFrq_k}(\sym_k) 
\nonumber \\ & \quad 
\times \prod_{l}\msg{\chTime_l}{\rSigFrq_k}(\chTime_l) \prod_{j\neq t}\msg{\impN_j}{\rSigFrq_k}(\impN_j) 
\end{align}
which poses the same difficulties as \eqref{eq:posterior} and \eqref{eq:msgRSigFrq2ch}.

Although GAMP can help approximate the messages in \eqref{eq:msgRSigFrq2noise}, GAMP alone is insufficient due to connections between the $\{d_t\}$ nodes, which are used to model the burstiness of the time-domain impulse-noise $\{i_t\}$.
However, recognizing that the underlying problem is estimation of a clustered-sparse sequence $\{\impN_t\}$ from compressed linear measurements, we can use the solution proposed in \cite{Schniter2010}, which alternated (G)AMP with the forward-backward algorithm \cite{Bishop2006}, as described below.

First, by temporarily treating the messages $\{\msg{d_t}{i_t}\}$, $\{\msg{\sym_k}{\rSigFrq_k}\}$, and $\{\msg{\chTime_l}{\rSigFrq_k}\}$ as fixed, we can apply $\ngamp[]$ under the likelihood model
\begin{equation}\label{eq:ngamplike}
\pdf{\rSigFrq_k|\impNFrq_k} \!= \!\!
\begin{cases}
\N{\rSigFrq_k;\impNFrq_k,\bckPwr} & \!\!\!\!\text{if } k\in\nullSet \\
\N{\rSigFrq_k;\pilot\hat{\chFrq}_k+\impNFrq_k,\pilotPwr\var_k^{\chFrq}\!+\!\bckPwr} & \!\!\!\!\text{if } k\in\pilotSet \\
\sum\limits_{l=1}^{\card{\constl}} \symBlf_k^{(l)} \N{\rSigFrq_k;\sym^{(l)}\hat{\chFrq}_k+\impNFrq_k,\pilotPwr\var_k^{\chFrq}\!+\!\bckPwr} & \!\!\!\!\text{if } k\in\dataSet
\end{cases}
\end{equation}
implied by \eqref{eq:subchannels} and \eqref{eq:ofdmFrqDomain}, 
and the coefficient prior
\begin{equation}\label{eq:impulsiveNoiseBlf}
\pdf{\impN_t} = \mixPrb_t^{(0)} \delta(\impN_t) + \sum_{k=1}^{\nMixComp-1} \mixPrb_t^{(k)} \N{\impN_t;0,\var^{(k)}}
\end{equation}
implied by \eqref{eq:impPdf}.
In \eqref{eq:ngamplike}, $\symBlf_k^{(i)}=\msg{\sym_k}{\rSigFrq_k}(\sym^{(i)})$ are the symbol beliefs coming from the $\{\sym_k\}$ nodes and $\{\hat{\chFrq}_k,\var_k^{\chFrq}\}$ are the frequency-domain channel estimates previously calculated by $\chgamp[]$.
Meanwhile, in \eqref{eq:impulsiveNoiseBlf}, $\{\mixPrb_t^{(k)}\}_{k=0}^{K-1}$ represents the pmf on the noise state $z_t$ that is set as $\mixPrb_t^{(k)}=\msg{\nState_t}{d_t}(k)/(\sum_{l=0}^{K-1} \msg{\nState_t}{d_t}(l))$.
The resulting output MMSE estimation functions, derived in Appendix~\ref{appdx:noiseDerivation}, are listed in \tablename~\ref{tab:outputScalarEstimationFun}, 
and the input MMSE estimation functions are 
\begin{align}
\E{\impN_t|\frqvc{\rSigFrq},\hat{r},\var^{r}} &= \sum\limits_{k=0}^{\nMixComp-1} \alpha_t^{(k)} \frac{\var^{(k)}\est{r}}{\var^{(k)}+\var^r}\\
\varOp{\impN_t|\frqvc{\rSigFrq},\hat{r},\var^{r}} 
&= \sum\limits_{k=0}^{\nMixComp-1}\frac{\alpha_t^{(k)}}{\var^r + \var^{(k)}}\left(\var^r\var^{(k)} + \frac{\abs{\var^{(k)}\hat{r}}^2}{\var^r + \var^{(k)}} \right) 
\nonumber\\&\quad 
-\abs{\E{\impN_t|\frqvc{\rSigFrq},\hat{r},\var^{r}}}^2 .
\end{align}
Here, $\{\alpha_t^{(k)}\}_{k=0}^{K-1}$ is the posterior pmf for noise-state $\nState_t$, with
\begin{equation}\label{eq:posStateBlf}
\alpha_t^{(k)} = \Prb{\nState_t\!=\!k|\est{r}}= \frac{\pdf{\est{r}|\nState_t=k}\mixPrb_t^{(k)}}{\sum_{l=0}^{\nMixComp-1} \pdf{\est{r}|\nState_t=l}\mixPrb_t^{(l)}}
\end{equation} 
where $\pdf{\est{r}|\nState_t\!=\!k}=\N{\est{r};0,\var^r+\var^{(k)}}$ is the noise state likelihood. 

Using these input and output MMSE estimation functions, $\ngamp[]$ is iterated until convergence, generating (for each $t$) an outgoing belief $\msg{\impN_t}{d_t}(i_t)= \N{\impN_t;\est{r},\var^r}$ about the noise-impulse $\impN_t$. 
This belief flows through the factor node $d_t$ which, according to the SPA, gives 
the rightward flowing noise-state belief 
\begin{equation}
\msg{d_t}{\nState_t}(\nState_t\!=\!k) \propto \N{\est{r};0,\var^r+\var^{(k)}} 
\end{equation} 
that acts as a prior for ``Markov-chain (MC) decoding,'' i.e., inference on the rightmost sub-graph in \figurename~\ref{fig:factorSystemModel}.
Since the MC sub-graph is non-loopy, it suffices to apply one pass of the forward-backward algorithm; see \cite{Bishop2006} for details. 
Subsequently the refined noise-state beliefs $\{\msg{z_t}{d_t}\}$ are passed back to the noise subgraph where each is used to compute the corresponding pmf $\{{\mixPrb}_t^{(k)}\}_{k=0}^{K-1}$ used in \eqref{eq:posStateBlf} by the next invokation of $\ngamp[]$.

When the noise-state beliefs $\{\msg{z_t}{d_t}\}$ have converged, the impulse-noise iterations are terminated and the $\{\est{\impNFrq}_k,\var_k^{\impNFrq}\}$ produced by $\ngamp[]$ are close approximations to the marginal posterior means and variances of $\{\impNFrq_k\}$ that will be used by $\chgamp[]$ in the next turbo iteration.
In addition, for each data tone $k\in\dataSet$, $\ngamp[]$ yields the leftward flowing soft symbol beliefs 
\begin{equation}
\label{eq:symbelief}
\msg{\rSigFrq_k}{\sym_k}(\sym) = \N{\rSigFrq_k;\sym\hat{\chFrq}_k+\hat{\impNFrq}_k,\abs{\sym}^2\var^{\chFrq}_k+\var^{\impNFrq}_k+\bckPwr}
\end{equation}
that are subsequently used for decoding (as described below).
Here, $\{\hat{\chFrq}_k,\var^{\chFrq}_k\}$ and $\{\hat{\impNFrq}_k,\var^{\impNFrq}_k\}$ play the role of soft frequency-domain channel and impulse-noise estimates, respectively.

Note that if the noise $\{i_t\}$ is \emph{not} modeled as bursty, then there is no need to apply the forward-backward algorithm and it suffices to run $\ngamp[]$ only once per turbo iteration.
In this case, \eqref{eq:impulsiveNoiseBlf} reduces to \eqref{eq:impPdf} and $\mixPrb_t^{(k)}$ reduces to the time-invariant prior parameter $\mixPrb^{(k)}$ discussed in Section~\ref{sec:ImpulsiveNoiseModels}. 

\subsubsection{Symbols to Bits}

The SPA dictates that the messages flowing leftward through the symbol nodes $\{\sym_k\}$ come out unchanged, i.e., $\msg{\sym_k}{\symMap_k}=\msg{\rSigFrq_k}{\sym_k}$.
Moreover, it dictates that the message flowing leftward out of the symbol-mapping node $\symMap_k$ and into the coded-bit node $\codedBit_{k,m}$ takes the form
\begin{align}
\msg{\symMap_k}{\codedBit_{k,m}}(\codedBit) &= \sum_{l=1}^{\card{\constl}}\sum_{\vc{\codedBit}_k\exc\codedBit_m} \Prb{\sym^{(l)}|\vc{\codedBit}_k} \msg{\sym_k}{\symMap_k}(\sym^{(l)}) 
\nonumber\\&\quad \times 
\prod_{m'\neq m}\msg{\codedBit_{k,m'}{\symMap_k}}(\codedBit_{m'}) \\
& = \frac{\sum_{l:\codedBit_m^{(l)}=c} \msg{\sym_k}{\symMap_k}(\sym^{(l)}) \msg{\symMap_k}{\sym_k}(\sym^{(l)})}{\msg{\codedBit_{k,m}}{\symMap_k}(\codedBit)} 
\end{align}
where the last step was is derived in \cite{Schniter2011}.

Finally, the computed coded-bit beliefs are passed to the coding/interleaving factor node. 
This can be viewed as passing (extrinsic) soft information into a soft-input/soft-output (SISO) decoder, where it is treated as prior information for decoding according to the ``turbo'' principle. 
SISO decoding has been studied extensively and we refer the interested reader to \cite{MacKay2003} for a detailed account. 
After SISO decoding terminates, it will produce extrinsic soft information, in the form of beliefs $\{\msg{\codedBit_{k,m}}{\symMap_k}\}$, that will be passed rightward to the symbol-mapping nodes at the start of the next turbo iteration. 
The turbo iterations are terminated after either the decoder detects no bit errors, the beliefs $\{\msg{\codedBit_{k,m}}{\symMap_k}\}$ have converged, or a maximum number of turbo iterations has elapsed.

\subsection{Simplified Receivers}
\label{sec:SimplifiedReceivers}
Although the JCISB receiver, as presented in Section~\ref{sec:JointChannelAndNoiseEstimationAndDecoding}, utilizes all data, pilot, and null tones to perform inference over the complete factor graph in \figurename~\ref{fig:factorSystemModel}, the proposed framework is flexible in that it can be easily modified to provided a desired trade-off between performance and computational complexity. 
For example, due to computational or architectural constraints, one might opt to simplify the receiver by either 1) using only a subset $\uTones$ of tones, or 2) replacing variable nodes in the factor graph with fixed exogenous soft estimates of those variables. 

Since reducing the size of the tone subset $\uTones$ will reduce both receiver complexity and performance (see Section~\ref{sec:TonesAllocationAndSelection}), the selection of $\uTones$ should be done carefully to balance these conflicting objectives.
In the sequel, we will denote the JCISB receiver that utilizes only the tone subset $\uTones\subset\{\dataSet\cup\pilotSet\cup\nullSet\}$ by $\JCISB{\uTones}$.
A generic implementation of $\JCISB{\uTones}$ would execute the steps in Section~\ref{sec:TonesAllocationAndSelection} but with $\chgamp$ and $\ngamp$, and then compute approximate-MMSE estimates of $\{\chFrq_k\}_{k\in\notuTones}$ and $\{\impNFrq_k\}_{k\in\notuTones}$ at $\notuTones=\{\dataSet\cup\pilotSet\cup\nullSet\}\setminus\uTones$ using GAMP's time-domain approximate-MMSE estimates $\{\est{\chTime}_t,\var^{\chTime}_t\}$ and $\{\est{\impN}_t,\var^{\impN}_t\}$ and the linear relationships $\frqvc{\chFrq}=\sqrt{N}\DFT\vc{\chTime}$ and $\frqvc{\impNFrq}=\DFT\vc{\impN}$.
That said, the case $\uTones=\pilotSet\cup\nullSet$ deserves special attention, since
here it suffices to perform joint channel and impulse (JCI) estimation in a manner that is decoupled from symbol and bit estimation. 

There are several ways that one might remove variable nodes from the factor graph in \figurename~\ref{fig:factorSystemModel} to simplify the resulting JCISB receiver (at the expense of performance: see Section~\ref{sec:NumericalResults}).  For example,
\subsubsection{Non-bursty JCISB}
Here the time-domain impulse-noise $\{\impN_t\}$ is modeled as non-bursty, in which case it suffices to remove the noise-state nodes $\{\nState_t\}$, use the GM prior \eqref{eq:impPdf} in the factor nodes $\{d_t\}$, and execute one impulse-noise iteration (without the forward-backward algorithm) per turbo iteration.
\subsubsection{Joint channel, symbol, and bit (JCSB) estimation}
Here we separately estimate $\{\impNFrq_k\}$ from only the null tones using $\ngamp[]$, and then fix the resulting soft estimates $\{\est{\impNFrq},\var^{\impNFrq}\}$ over the turbo iterations, avoiding the need to run $\ngamp[]$ more than once.
\subsubsection{Joint impulse, symbol, and bit (JISB) estimation}
Here we compute soft linear-MMSE estimates of the frequency-domain channel coefficients $\{\chFrq_k\}$ and use these in place of the GAMP-computed nonlinear-MMSE estimates $\{\est{\chFrq}_k,\var^{\chFrq}_k\}$, avoiding the need to ever run $\chgamp[]$. 
\subsubsection{GAMP-impulse (GI) estimation}
Here we first LMMSE estimate $\{\chFrq_k\}$ from the pilot tones, then use those outputs with $\ngamp[]$ to estimate $\{\impNFrq_k\}$ from the pilot and null 
tones, and finally use both the soft channel and impulse estimates to recover the symbols and bits via standard SISO decoding.
The principal feature distinguishing this approach from conventional OFDM estimation is the use of GAMP-impulse estimation from pilot and null tones. 
The GI provides an important reference point since it uses the same information provided by the null and pilot tones as the prior work in \cite{Caire2008,Lampe2011,Lin2012}.

\subsection{Computational Complexity}
\label{sec:ComputationalComplexity}
The computational complexity of JCISB stems primarily from repeated calls to the GAMP algorithm, whose complexity grows as $O(N\log N + N|\constl|)$ per GAMP iteration. 
For small constellations $\constl$, GAMP's per-iteration complexity is dominated by steps (R2) and (R8) of \tablename~\ref{tab:gamp}, which can each be implemented in $\frac{N}{2}\log_2 N$ multiplies using an $N$-length FFT; 
due to the constant-modulus nature of the entries of the DFT-matrix $\vc{\Phi}$, 
steps (R1) and (R7) reduce to simple summations.
For large constellations like $1024$-QAM, steps (R3)--(R4), which involve $\nDataTones$ summations of $|\constl|$ terms each (see \tablename~\ref{tab:outputScalarEstimationFun}), may also be of significant complexity.

As discussed, the simplifications discussed in Section~\ref{sec:SimplifiedReceivers} can be used to reduce the complexity.
For example, when $\uTones$ does not include data tones, the GAMP likelihoods \eqref{eq:chgamplike} and \eqref{eq:ngamplike} do not involve the $|\constl|$-term summations and so GAMP complexity is no longer dependent on $|\constl|$.
Even in that case, though, the final $\nDataTones$ symbol beliefs \eqref{eq:symbelief} must be evaluated at all $|\constl|$ symbol possibilities $\sym\in\constl$, and so the proposed receiver complexity order remains at $O(N\log N + N|\constl|)$, as does that of the conventional OFDM receiver.

In contrast, the state-of-the-art approach \cite{Lin2012} uses the SBL algorithm for impulse-noise estimation, an iterative approach that computes a matrix inversion at each iteration, and thus has overall complexity $O(N^3 + N|\constl|)$.
Thus, given that $N$ is usually in the hundreds or thousands, the proposed JCISB approach will require much less computation than \cite{Lin2012}.

\subsection{Pilot and Null Tone Placement and Selection}
\label{sec:TonesAllocationAndSelection}

In conventional OFDM systems, it is typical to place pilot tones on a uniformly spaced grid, as this yields MMSE optimal channel estimates in AWGN-corrupted frequency-selective channels \cite{Negi1998}. 
Meanwhile, it is customary to place null tones at the spectrum edges in order to reduce out-of-band emissions \cite{prime}. 
These practices, however, should be re-examined when the receiver is expected to operate in the presence of impulsive noise, since there the MMSE channel estimator is nonlinear and the frequency-domain noise is dependent across tones, making it suboptimal to ignore null-tones while decoding.

Viewing impulse-noise estimation as a sparse reconstruction problem \cite{Caire2008}, we realize that the placement $\uTones$ of the tones used for estimation strongly affects the isometry of the linear transformation $\DFT_{\uTones}$ relating the sparse tone sequence $\vc{\impN}$ to the linearly compressed measurements $\frqvc{\rSigFrq}_{\uTones}$.
For sparse signal reconstruction, recovery guarantees can be stated when the measurement matrix $\grkvc{\Phi}$ has sufficiently low \emph{coherence} \cite{Studer2012}
\begin{equation}\label{eq:coherenceDef}
\coh(\grkvc{\Phi}) = \max_{k,l,k\neq l} \frac{\abs{\conj{\grkvc{\phi}}_k \grkvc{\phi}_l}}{\|\grkvc{\phi}_k\|_2 \|\grkvc{\phi}_l\|_2}
\end{equation}
using $\grkvc{\phi}_k$ to denote the $k$th column of $\grkvc{\Phi}$.
Section~\ref{sec:NumericalResults} provides evidence that $\coh(\DFT_{\uTones})$ predicts the performance of tone placement $\uTones$ in impulse-noise corrupted OFDM.

\section{Numerical Results}
\label{sec:NumericalResults}

In this section, we evaluate the performance of our proposed JCISB receivers using Monte-Carlo simulations, comparing to both existing work and fundamental bounds. 
We demonstrate that, in both coded and uncoded scenarios, the proposed JCISB framework provides significant performance gains over existing techniques at a computational complexity only slightly higher than the conventional DFT receiver and thus significantly lower than the best performing prior work. 
In fact, we show that JCISB performs within $1$dB of theoretical performance bounds, establishing its near-optimality.
Furthermore, we conduct numerical studies that investigate the impact of 
receiver simplifications, 
impulse-noise modeling and mitigation, 
and pilot/null tone placement.

\subsection{Setup}
\label{sec:Setup}

Unless stated otherwise, pilot tones were spaced on a uniform grid while the null tones were placed randomly. 
Noise realizations were generated according to one of the two models described in Section~\ref{sec:ImpulsiveNoiseModels}: non-bursty i.i.d-GM noise, having two impulsive noise states with powers $20$dB and $30$dB above the background noise occurring $7\%$ and $3\%$ or the time, respectively; and bursty GHMM noise, with the same marginal statistics but with temporal dynamics governed by the state transition matrix in \eqref{eq:exampleTrsMatrix}.
Unless noted otherwise, JCISB was run using at most $5$ turbo iterations, $5$ noise iterations, and $15$ GAMP iterations. 
Throughout, signal-to-noise ratio ($\SNR$) refers to the ratio of the received signal power to the total noise power.

\subsection{Comparison with Existing Schemes}
\label{sec:ComparisonWithOtherSchemes}
\begin{figure}[!t]
\centering
\includegraphics[width=2.75in]{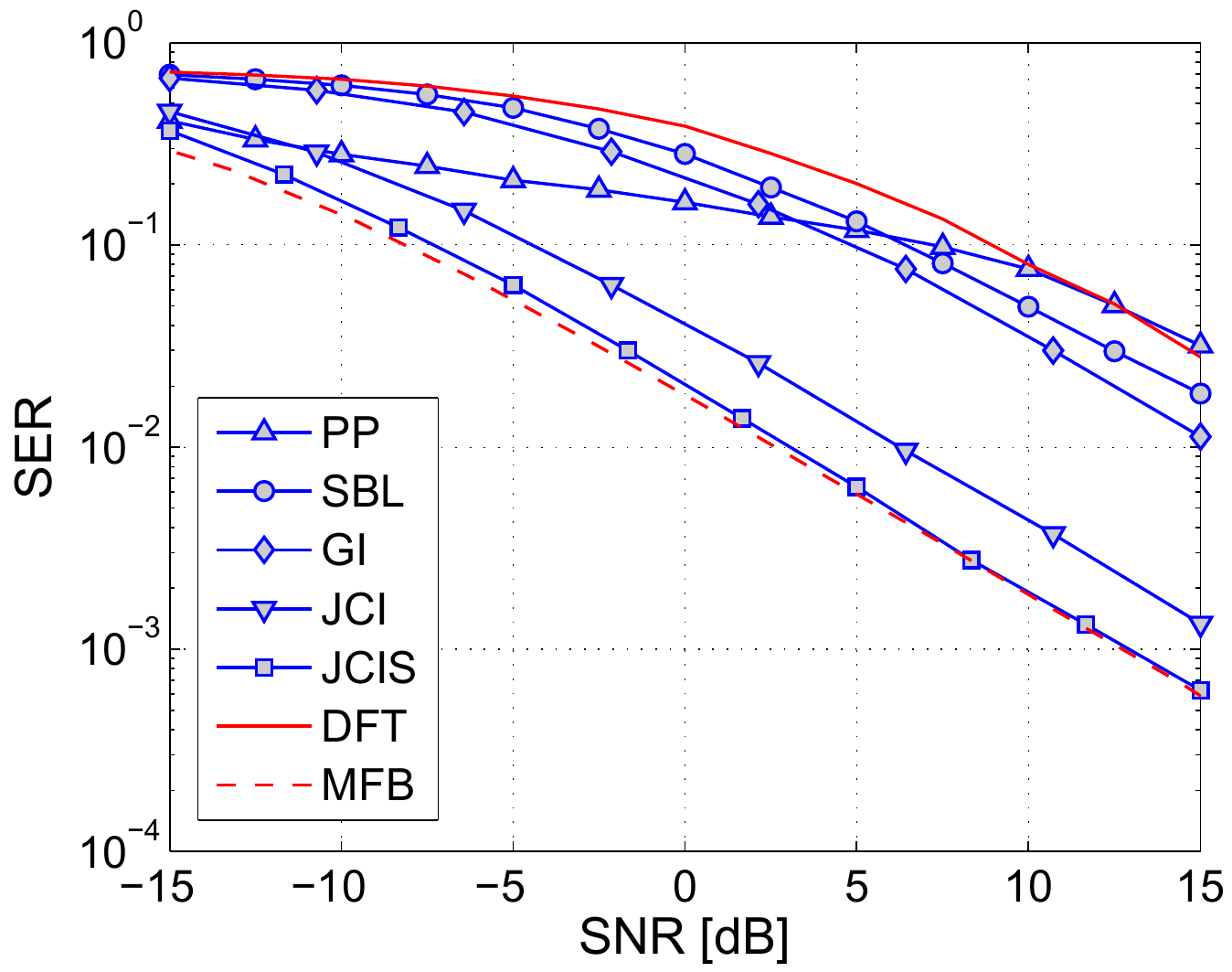}
\caption{Uncoded \SER versus \SNR for $4$-QAM OFDM with $256$ total tones, $80$ null tones, and $15$ pilot tones over a $5$-tap Rayleigh-fading channel in i.i.d GM noise.
}
\label{fig:comparisonWithOtherSchemes}
\end{figure}

\figurename~\ref{fig:comparisonWithOtherSchemes} plots uncoded symbol-error rate ($\SER$) versus \SNR for a prototypical PLC setting: $4$-QAM modulated OFDM with $256$ subcarriers, of which $80$ tones are nulls and $15$ are pilots, under a $5$-tap Rayleigh-fading channel corrupted by i.i.d GM noise. 
In \figurename~\ref{fig:comparisonWithOtherSchemes},
the ``JCIS'' trace represents our proposed JCISB approach but without bit estimation (since here we evaluate uncoded $\SER$), and the ``GI'' trace represents the proposed GI simplification from Section~\ref{sec:SimplifiedReceivers}.
The ``DFT'' trace represents the conventional OFDM receiver, which performs LMMSE pilot-aided channel estimation, LMMSE equalization, and decoupled symbol-detection on each equalized tone.
The ``PP'' trace refers to \cite{Haring2001}, which performs MMSE-optimal processing prior to conventional OFDM reception and has been shown to perform best among the ``pre-processing'' techniques discussed in Section~\ref{sec:PriorWork}. 
The ``SBL'' trace refers to \cite{Lin2012}, which was recently shown to perform best among the ``sparse reconstruction'' methods.
Here, the PP and SBL approaches include LMMSE channel estimation, whereas in the original formulations \cite{Haring2001,Lin2012} the channel was treated as known.
The ``MFB'' trace shows the matched-filter bound, which computes tone-averaged $\SER$ assuming that each symbol is detected under perfect knowledge of every other symbol as well as the channel.
By subtracting the known effect of the other symbols, the received signal under MFB is given by
\begin{equation} \label{eq:MFBound}
\vc{\rSig} = \mathbf{H} \IDFT (\sym\stdBasis_k ) + \vc{\nTime} = \sym\vc{\bar{f}}_k  +  \vc{\nTime}
\end{equation}
where the unknown symbol $\sym$ is sent on tone $k$ and where  $\stdBasis_k$ is the standard basis and $\vc{\bar{f}}_k$ is the $k$-th column of $\mathbf{H}\IDFT$. Using the factorization of the noise pdf in time domain, it is straightforward to find the MAP detection rule for $\sym$ \cite{Spaulding1977}.
Due to the non-Gaussianity of the noise, we evaluated the MFB via Monte-Carlo.

The \SER curves in \figurename~\ref{fig:comparisonWithOtherSchemes} show that the proposed JCIS receiver drastically outperforms the conventional OFDM receiver (by $15$dB), 
the PP receiver (by $15$dB in the high \SNR regime), 
and the state-of-the-art SBL receiver (by $13$dB). 
We attribute these huge performance gains to the fact that JCIS utilizes \emph{all} received tones (pilots, nulls, and data) for \emph{joint} channel, impulse, and symbol estimation.
In contrast, PP does not use OFDM signal structure for impulse-noise mitigation; and SBL decouples the estimation of the channel, impulses, and symbols, and performs linear MMSE channel estimation using only pilot tones, which not only ignores information on data and null tones, but is also strongly suboptimal in the presence of impulsive noise.
Moreover, the proposed JCIS receiver follows the MF bound to within $1$dB over the full SNR range, demonstrating its near-optimality.
\figurename~\ref{fig:comparisonWithOtherSchemes} also shows that the proposed JCI simplification performs only $3$dB worse than JCIS, and that the GI simplification performs $13$dB worse than JCIS but $2$dB better than the state-of-the-art\footnote{Although PP outperforms both SBL and GI when $\SNR<3$, the achieved $\SER$s are unusably high.} SBL receiver.

\subsection{Impact of Impulse-Noise Modeling and Mitigation}
\label{sec:TheValueOfImpulsiveNoiseModeling}

In this section, we evaluate the relative success of various strategies for modeling and mitigating impulsive noise in OFDM, again restricting our attention to uncoded transmissions.
For clarity, we consider a trivial (unit-gain non-fading) channel that is perfectly known to the receiver, and thus we include no pilot tones.
Without channel estimation and bit decoding, our proposed JCISB approach then reduces to JIS.
Below, we compare JIS to the SBL receiver \cite{Lin2012} and to the GI simplification proposed in Section~\ref{sec:SimplifiedReceivers}.
Given the absence of pilot tones, GI and SBL are quite similar: both perform impulse-noise estimation using only null tones and in a manner that is decoupled from symbol estimation.

We first compare the noise-estimation performance of JIS, GI, and SBL using the normalized mean squared estimation error metric
$\NMSE
=\E{|\nTime_t-\est{\impN}_t|^2}/\E{|\nTime_t|^2}
=\E{|\nFrq_k-\est{\impNFrq}_k|^2}/\E{|\nFrq_k|^2}
$, which can be interpreted as follows.
Recalling that \SER increases proportionally to $\MSE=\E{|\nFrq_k-\est{\impNFrq}_k|^2}=\E{|\bckNFrq_k+(\impNFrq_k-\est{\impNFrq}_k)|^2}$, which includes both background noise and impulse-estimation error, and noticing\footnote{For a unit signal power, $\SNR^{-1}=\E{|\nFrq_k|^2}$, so that $\NMSE=\MSE\times\SNR$ and thus $\MSE^{-1}=\SNR/\NMSE$.} 
that $\MSE^{-1}=\SNR/\NMSE$, we recognize $\NMSE$ as the factor by which the \emph{effective} signal-to-noise ratio $\MSE^{-1}$ is smaller than the \emph{stated} signal-to-noise ratio $\SNR$.

\begin{figure}[!t]
\centering
\includegraphics[width=3.52in]{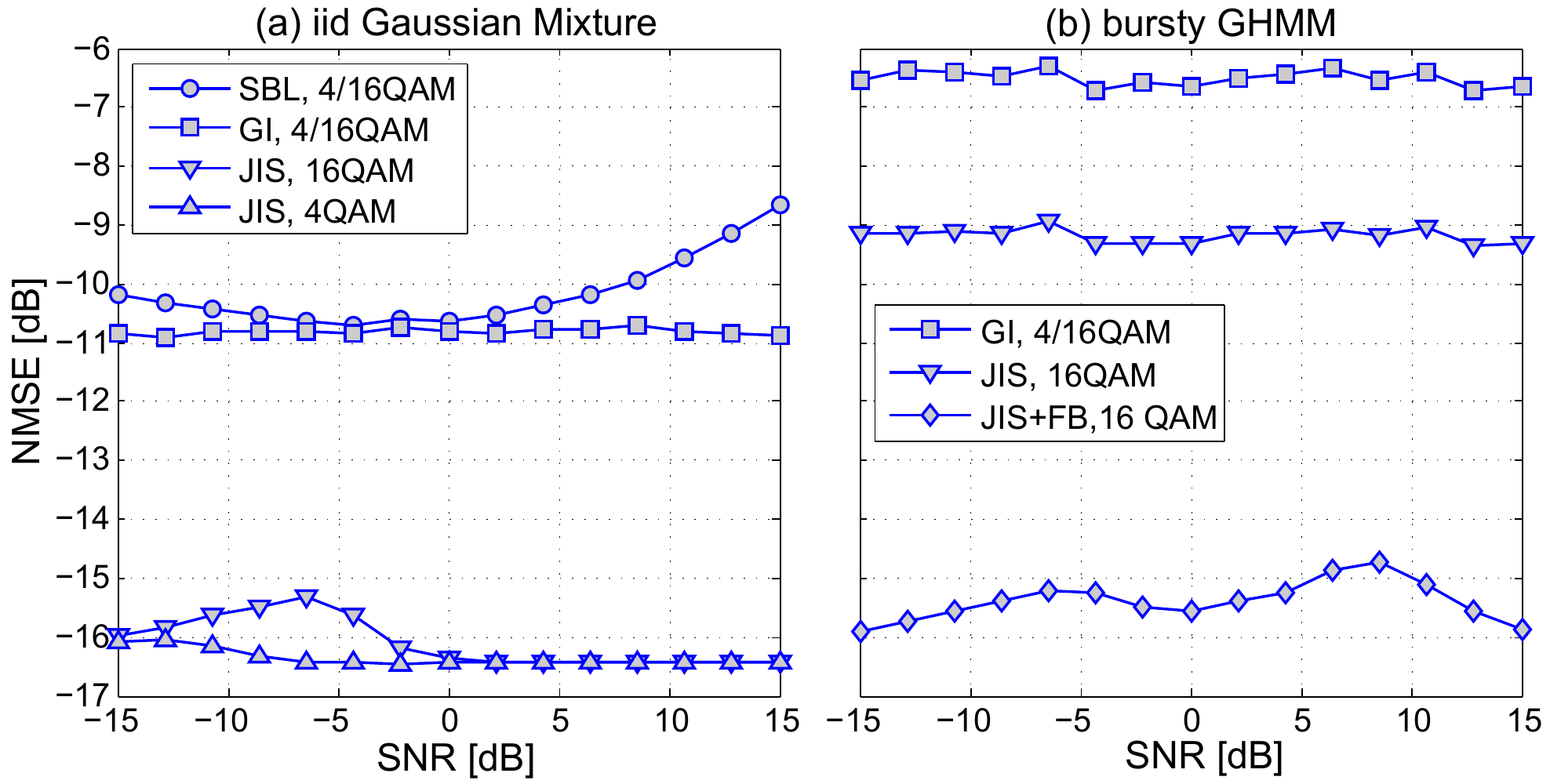}
\caption{\NMSE versus \SNR when estimating the noise sequence $\vc{\nTime}$ for (a) i.i.d-GM and (b) GHMM noise models, for OFDM with $256$ total tones and $60$ null tones, under a known trivial channel.  
The corresponding \SER performance is plotted in \figurename~\ref{fig:ValueOfModeling}.
}
\label{fig:noiseEstimateNMSE}
\end{figure}

Figure~\ref{fig:noiseEstimateNMSE}(a) plots \NMSE in the estimation of i.i.d GM noise versus \SNR for the JIS, GI, and SBL receivers.
The GI traces in \figurename~\ref{fig:noiseEstimateNMSE}(a) imply that GAMP is a uniformly better estimator of i.i.d GM noise than SBL, although the difference is $<1$dB for $\SNR$s between $-15$ and $8$ dB.
This behavior is expected, given that the underlying problem is one of estimating a length-$256$ i.i.d-GM sequence from $60$ randomly selected Fourier observations, for which the superiority of GM-GAMP over SBL was established in \cite{Vila2012}.
The JIS traces in \figurename~\ref{fig:noiseEstimateNMSE}(a) show $\NMSE$s that are significantly (i.e., $\geq 5$dB) better than GI and SBL across the $\SNR$ range, and this is because JIS uses both null and data tones, rather than just null tones.
To extract meaningful noise information from the data tones, JIS must accurately infer the data symbols. 
The latter is easier with 4-QAM than with 16-QAM, which explains the gap between the traces in \figurename~\ref{fig:noiseEstimateNMSE}(a).

Figure~\ref{fig:noiseEstimateNMSE}(b) plots \NMSE in the estimation of GHMM noise versus \SNR for the proposed JIS receiver with the forward-backward (FB) iterations, and two simplifications: JIS without FB (labelled as ``JIS'' for consistency with \figurename~\ref{fig:noiseEstimateNMSE}(a)) and GI.
Comparing \figurename~\ref{fig:noiseEstimateNMSE}(b) to \figurename~\ref{fig:noiseEstimateNMSE}(a), we see that GHMM noise is significant more challenging than i.i.d-GM noise: the \NMSE of JIS is $7$dB worse, and that of GI is $5$dB worse, in the GHMM case
However, the FB iterations help significantly: they restore approximately $6$dB of the lost \NMSE. 

\begin{figure}[!t]
\centering
\includegraphics[width=3.5in]{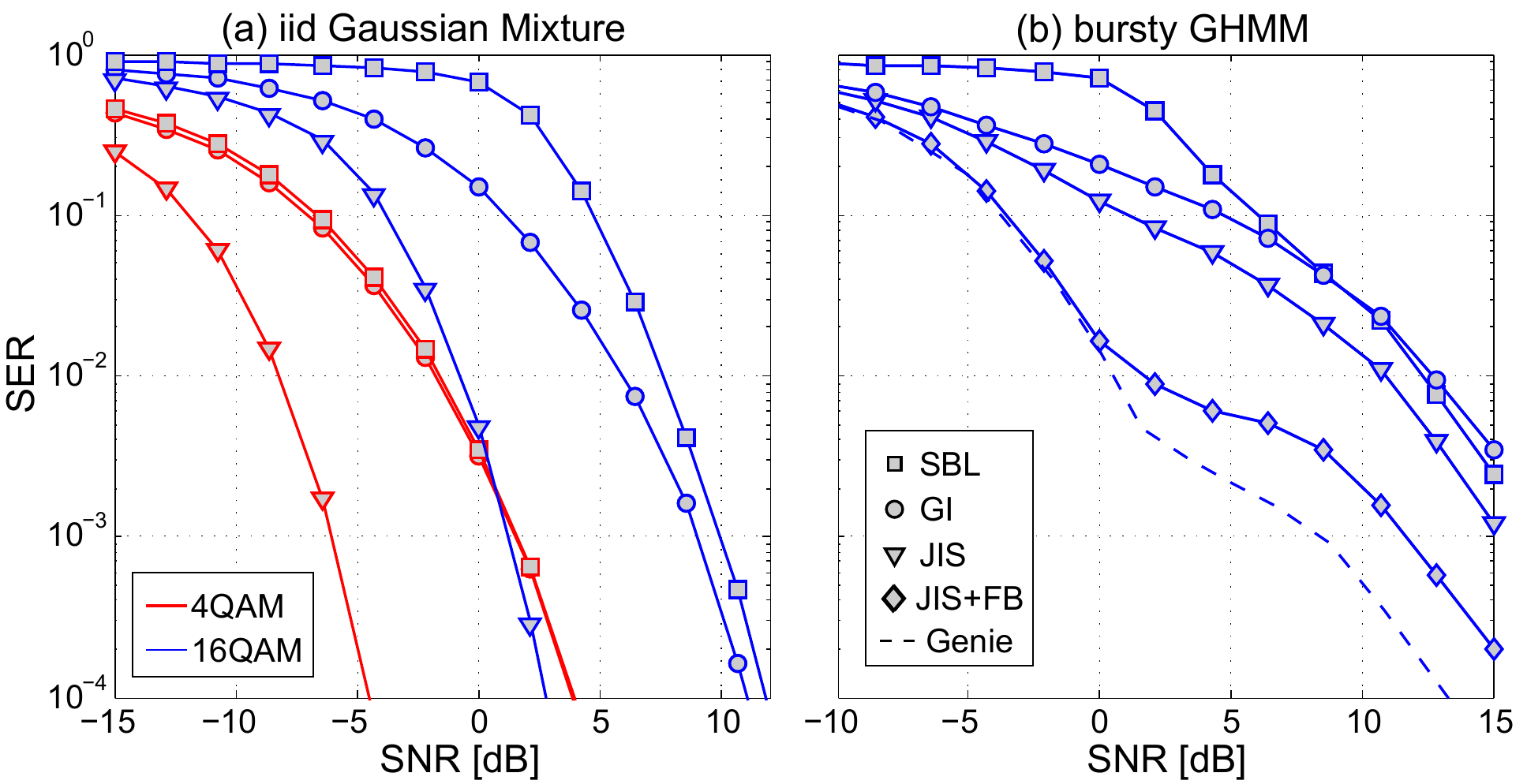}
\caption{Uncoded \SER versus \SNR for OFDM with $256$ total tones and $60$ null tones under a known trivial channel for (a) i.i.d GM and (b) GHMM noise.  Red traces denote 4-QAM while blue traces denote 16-QAM.
}
\label{fig:ValueOfModeling}
\end{figure}

Next we compare the \SER performance of JIS, GI, and SBL in the same trivial-channel setting. 
Figure~\ref{fig:ValueOfModeling}(a) shows the case of i.i.d-GM noise.
There we see that JIS significantly outperforming SBL with both 4-QAM (red) and 16-QAM (blue) constellations, as expected from the superior noise-estimation \NMSE in \figurename~\ref{fig:noiseEstimateNMSE} and from the fact that JIS estimates the symbols \emph{jointly} with the noise impulses.  
Meanwhile, it shows GI performing on par with SBL with 4-QAM but somewhat better than SBL with 16-QAM, especially at medium $\SNR$.

Figure~\ref{fig:ValueOfModeling}(b) then shows \SER under GHMM (i.e., bursty) noise.
Comparing \figurename~\ref{fig:ValueOfModeling}(b) to \figurename~\ref{fig:ValueOfModeling}(a), we see that the burstiness of the noise causes the \SER of \emph{all} receivers to degrade significantly.
Moreover, this degradation persists when the JIS receiver uses MC iterations, even though the \NMSE results in \figurename~\ref{fig:noiseEstimateNMSE} show only about a $1$dB loss due to burstiness.
We attribute the \SER sensitivity to the fact that the noise burstiness makes some OFDM-symbols much more noise-corrupted than others, and those heavily corrupted symbols skew the \emph{average} $\SER$ reported in \figurename~\ref{fig:ValueOfModeling}(b).
Regardless, \figurename~\ref{fig:ValueOfModeling}(b) shows that the FB-assisted JIS receiver significantly outperforms non-FB-assisted JIS, GI, and the state-of-the-art SBL algorithm, especially at medium $\SNR$.
To investigate whether the kink in the JIS$+$FB trace was due to suboptimality of the FB noise-state inference, we simulated a genie-aided receiver that knows the true state of the GHMM noise at each time index. 
Since the genie trace also exhibits the kink, it is evidently not due to suboptimality of FB.

\subsection{Impact of Pilot and Null Tone Placement}
\label{sec:KnownToneAllocation}

\begin{figure}[!t]
\centering
\includegraphics[width=2.75in]{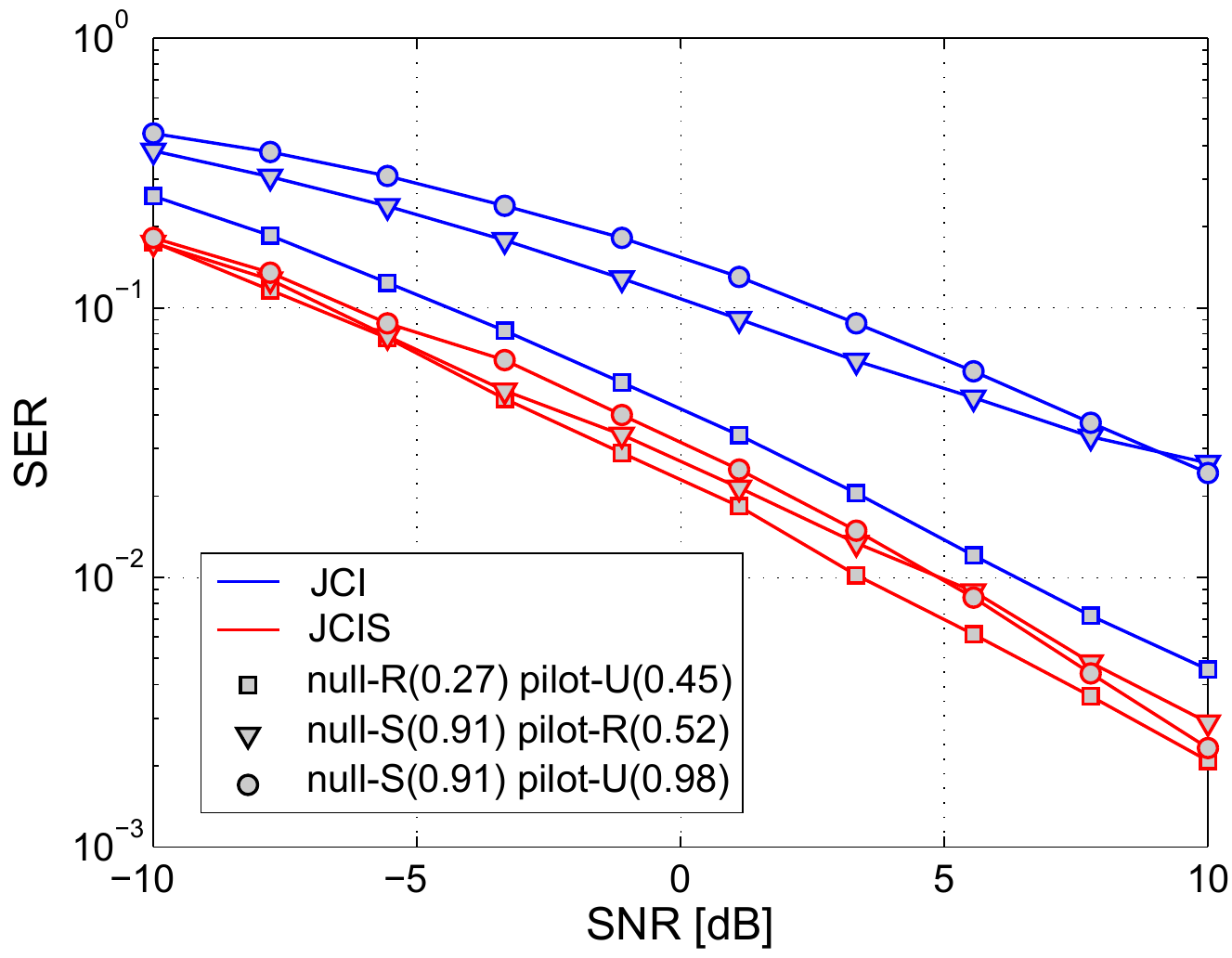}
\caption{Uncoded \SER for OFDM with $256$ total tones, $60$ null tones, and $25$ pilot tones under a $5$-tap Rayleigh-fading channel and i.i.d-GM noise.  
Several null and pilot tone placements are considered: random (R), uniform (U), and sideband (S), with the corresponding coherence [$\coh(\DFT_\nullSet)$ or $\coh(\DFT_\pilotSet)$, recall \eqref{eq:coherenceDef}] specified in the legend.
}
\label{fig:knownToneAllocation}
\end{figure}

In this section, we investigate the impact of pilot and null tone placement.
For this, we examine the uncoded \SER of a 4-QAM $256$-tone OFDM system under a $5$-tap Rayleigh-fading channel in i.i.d-GM noise for both the proposed JCIS receiver and its JCI simplification, the latter of which ignores data tones during channel and impulse-noise estimation.
\figurename~\ref{fig:knownToneAllocation} shows that the conventional placement of sideband null tones and uniform pilot tones produces the worst \SER performance. 
Randomizing the pilot locations alone provides a modest performance gain for both JCI and JCIS, while randomizing the null locations alone improves the \SER performance dramatically, especially for JCI.\footnote{We expect JCI to be more sensitive to null/pilot-tone placement than JCIS, since the former observes the channel and noise impulses only through those tones.}
We conjectured in Section~\ref{sec:TonesAllocationAndSelection} that the performance improvement observed with randomized pilot and null tone placements can be explained by the corresponding reduction in coherence $\coh(\DFT_\pilotSet)$ and $\coh(\DFT_\nullSet)$, and the coherence values reported in \figurename~\ref{fig:knownToneAllocation} lend credence to this conjecture. 

\subsection{Coded Systems}
\label{sec:CodedSystems}

Finally, we investigate the bit error rate ($\BER$) performance of JCISB in the coded scenario. 
For this, we used an LDPC-coded 16-QAM $1024$-tone OFDM system with $150$ pilot and $0$ null tones under a $10$-tap Rayleigh-fading channel and i.i.d-GM noise.  
The LDPC codes had code-word length $\approx 60\: 000$ and rate $1/2$, with a modified coder/decoder implementations from \cite{Kozintsev:SW}.
We also investigate the conventional OFDM receiver (``DFT'') as well as the JCIS simplification, which omits SISO decoding from the turbo iterations.
For both JCIS and DFT, we performed SISO decoding as the final step.
For all receivers, the maximum number of LDPC iterations was $50$. 

\begin{figure}[!t]
\centering
\includegraphics[width=2.75in]{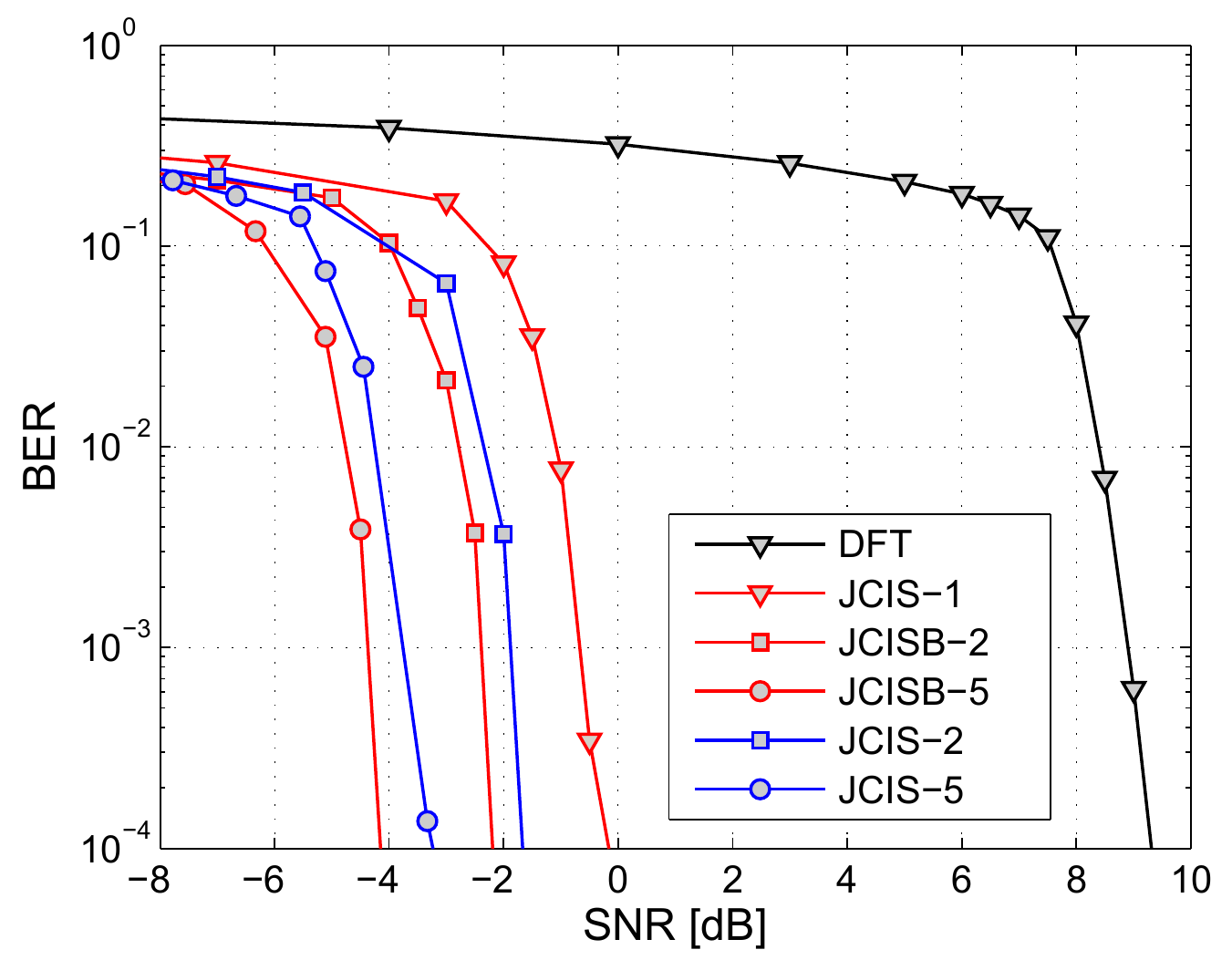}
\caption{Coded \BER for OFDM with $1024$ total tones, $150$ pilot tones, and $0$ null tones under a $10$-tap Rayleigh-fading channel in i.i.d-GM noise.
Here, JCISB-\# corresponds to JCISB after \# turbo iterations, and similar for JCIS-\#.
}
\label{fig:codedSystemBER}
\end{figure}

\figurename~\ref{fig:codedSystemBER} shows that, after only one turbo iteration, the proposed JCISB\footnote{Note that, with only a single turbo iteration, JCISB and JCIS are equivalent.} outperforms the conventional OFDM receiver by $10$dB.
Additional turbo iterations result in further gains of $4$dB.
\figurename~\ref{fig:codedSystemBER} also shows that JCIS's decoupling of bit estimation from channel, impulse, and symbol estimation costs approximately $1$dB.

\section{Conclusion }
\label{sec:Conclusion}

In a this paper, we presented a factor-graph approach to OFDM reception in multipath distorted and impulse-noise corrupted channels that performs near-optimal joint channel, impulse-noise, symbol, and bit (JCISB) estimation.
Our approach merges recent work on
generalized approximate message passing (GAMP) \cite{Rangan2011b}, its ``turbo'' extension to larger factor graphs \cite{Schniter2010}, and soft-input-soft-output SISO decoding \cite{MacKay2003}. 
Extensive numerical simulations show that the proposed JCISB receiver provides drastic performance gains over existing receivers for OFDM in impulsive noise, and performs within $1$dB of the matched-filter bound, all while matching the complexity order of the conventional OFDM receiver.
Furthermore, JCISB is easily parallelized, providing a natural mapping to FPGA implementations (see \cite{Nieman2013} for a recent FPGA implementation of the GI receiver).
Additional numerical simulations investigated the impact of JCISB simplifications, noise modeling and mitigation, and null/pilot tone placement.

\appendices

\section{Generalized Approximate Message Passing (GAMP)}\label{appndx:GAMPsummary}

\begin{figure}[!h]
\centering
\includegraphics[width=3.2in]{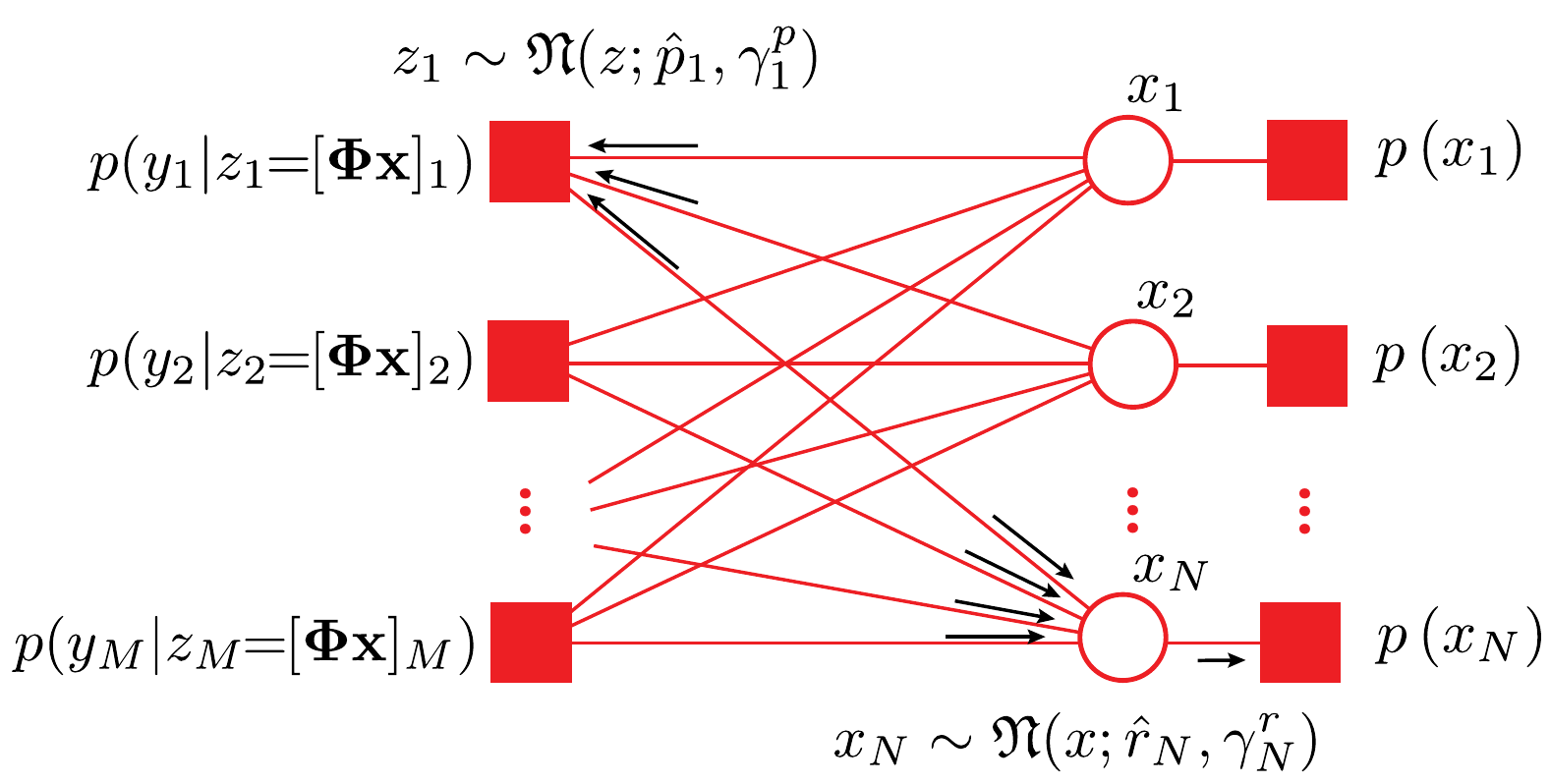}
\caption{Factor graph used to derive GAMP with illustrations of message approximations.}
\label{fig:gampFactorGraph}
\end{figure}

The GAMP algorithm, as developed in \cite{Rangan2011b} as an extension of AMP from \cite{Donoho2009}, addresses the estimation of a vector of independent possibly-non-Gaussian random variables $\vc{x}$ that are linearly mixed via linear transform $\grkvc{\Phi} \in \C^{M\times N}$ to form $\vc{z}=\grkvc{\Phi} \vc{x}=\trsp{[z_1 \cdots z_M]}$ and subsequently observed as $\vc{y}= \trsp{[y_1 \cdots y_M]}$ according to the likelihood function $\pdf{\vc{y}|\vc{z}}=\prod_{i=1}^M \pdf{y_i|z_i}$.
The GAMP algorithm is intended for the case when the dimensions $M$ and $N$ are both large, in which case the central limit theorem suggests approximating the product of messages flowing leftward into each factor node $f_i=\pdf{y_i|z_i}$ in \figurename~\ref{fig:gampFactorGraph} as $\prod_{j} \msg{x_j}{f_i}(z_i) \approx \N{z_i;\hat{p}_i,\var_i^p}$ and the product of messages flowing rightward into each variable node $x_j$ as 
$\prod_i \msg{f_i}{x_j}(x_j) \approx \N{x_j;\hat{r}_j,\var_j^r}$, where the quantities $\hat{p}_i$, $\var_i^p$, $\hat{r}_j$, and $\var_j^r$ can be computed from Table~\ref{tab:gamp}. Similarly, each outgoing message leaving a factor or variable node is approximated using a second order Taylor series expansion by two parameters.
The corresponding ``$\gamp{\vc{y},\vc{z},\grkvc{\Phi},\vc{x}}$'' algorithm is summarized in Table~\ref{tab:gamp}.
The detailed derivation and theoretical guarantees of the GAMP algorithm are beyond the scope of this paper; we refer the interested reader to \cite{Rangan2011b} and \cite{Javanmard2012} for more information. 

\section{Derivation of $\chgamp[]$ Likelihood}
\label{appdx:channelDerivation}
This appendix derives the output MMSE estimation functions $\E{\chFrq_k|\rSigFrq_k;\hat{p},\var^{p}}$ and $\varOp{\chFrq_k|\rSigFrq_k;\hat{p},\var^{p}}$ used in steps (R3)--(R4) of $\chgamp[]$.
We start with the case of data tones $k\in\dataSet$, where 
\begin{equation*}
\rSigFrq_k = \sym_k\chFrq_k  + \impNFrq_k + \bckNFrq_k .
\end{equation*}
$\rSigFrq_k$ is Gaussian when conditioned on $\sym_k$, and so according to the definition (D1) in Table~\ref{tab:gamp} and \cite{Kay1993},
\begin{equation}\label{eq:JCNEDDataChEgvS}
\E{\chFrq_k|\sym_k,\rSigFrq_k;\hat{p},\var^{p}} = \hat{p}+\frac{\conj{\sym}_k\var^p (\rSigFrq_k - \hat{\impNFrq}_k-\sym_k\est{p})}{\bckPwr + \var^{\impNFrq}_k+\abs{\sym_k}^2\var^{p}}.
\end{equation}
Given the belief $\{\vc{\symBlf}_k^{(l)}\}_{l=1}^{\card{\constl}}$ about symbol $\sym_k$, the \textit{law of total expectation} implies 
\begin{align}
\E{\chFrq_k|\rSigFrq_k;\hat{p},\var^{p}} &=\E[\sym_k|\rSigFrq_k]{\E{\chFrq_k|\sym_k,\rSigFrq_k;\hat{p},\var^{p}}}\\
&= \hat{p}+\sum\limits_{l=1}^{\card{\constl}}\posSymBlf_k^{(l)}\frac{\var^\impNFrq (\rSigFrq_k - \hat{\impNFrq}_k-\hat{p}\sym^{(l)})}{\bckPwr + \var^{\impNFrq}_k+\abs{\sym^{(l)}}^2\var^{p}}\label{eq:JCNEDDataChE}
\end{align}
where 
\begin{equation}\label{eq:postSym}
\posSymBlf_k^{(l)}=\Prb{\sym_k=\sym^{(l)}|\rSigFrq_k;\est{p},\var^p}\propto \pdf{\rSigFrq_k|\sym_k;\est{p},\var^p}\symBlf_k^{(l)}
\end{equation}
is the posterior symbol probability 
and $\pdf{\rSigFrq_k|\sym_k;\est{p},\var^p}=\N{\rSigFrq_k;\hat{\impNFrq}_k+\est{p}\sym_k,\var^{\impNFrq}+\abs{\sym_k}^2\var^{p}+\bckPwr}$.
Similarly, the \textit{law of total variance} implies
\begin{align}
\lefteqn{ \varOp{\chFrq_k|\rSigFrq_k;\hat{p},\var^{p}} 
	= \E[\sym_k|\rSigFrq_k]{\varOp{\chFrq_k|\sym_k,\rSigFrq_k;\hat{p},\var^{p}}} 
}
\nonumber\\&\hspace{30mm}
+ \varOp[\sym_k|\rSigFrq_k]{\E{\chFrq_k|\sym_k,\rSigFrq_k;\hat{p},\var^{p}}} \\
&= \sum\limits_{l=1}^{\card{\constl}}\posSymBlf_k^{(l)} 
\Bigg[\frac{\var^{p}\left(\bckPwr+\var_k^{\impNFrq}\right)}{\bckPwr + \var^{\impNFrq}_k+\abs{\sym^{(l)}}^2\var^{p}}  
+ \big|\E{\chFrq_k|\sym_k,\rSigFrq_k;\hat{p},\var^{p}}\big|^2
\Bigg] 
\nonumber\\ & \qquad 
- \big|\E{\chFrq_k|\rSigFrq_k;\hat{p},\var^{p}} \big|^2 .
\end{align}

The derivation for pilot tones $k\in \pilotSet$ reduces to the above under $\posSymBlf_k^{(1)}=1$, $\posSymBlf_k^{(l\neq 1)}=0$, and $\sym^{(1)}=\pilot$.

\section{Derivation of $\ngamp[]$ Likelihood}
\label{appdx:noiseDerivation}
This appendix derives the output MMSE estimation functions $\E{\impNFrq_k|\rSigFrq_k;\hat{p},\var^{p}}$ and $\varOp{\impNFrq_k|\rSigFrq_k;\hat{p},\var^{p}}$ used in steps (R9)--(R10) of $\ngamp[]$.
We start with the case of data tones $k\in\dataSet$, where
\begin{equation*}
\rSigFrq_k = \impNFrq_k + \sym_k\chFrq_k + \bckNFrq_k .
\end{equation*}
$\rSigFrq_k$ is Gaussian when conditioned on $\sym_k$, and so according to the definition (D1) in Table~\ref{tab:gamp} and \cite{Kay1993},
\begin{equation}\label{eq:JNEDDataChEgvS}
\E{\impNFrq_k|\sym_k,\rSigFrq_k;\hat{p},\var^{p}} = \hat{p} + \frac{\var^p(\rSigFrq_k - \hat{p}-\est{\chFrq}_k\sym_k)}{\bckPwr + \var^{p}+\abs{\sym_k}^2\var_k^{\chFrq}}.
\end{equation} 
Given the belief $\{\vc{\symBlf}_k^{(l)}\}_{l=1}^{\card{\constl}}$ about symbol $\sym_k$, the \textit{law of total expectation} implies
\begin{align}
\E{\impNFrq_k|\rSigFrq_k;\hat{p},\var^{p}} &=\E[\sym_k|\rSigFrq_k]{\E{\impNFrq_k|\sym_k,\rSigFrq_k;\hat{p},\var^{p}}}\\
&= \hat{p}+\sum\limits_{l=1}^{\card{\constl}}\posSymBlf_k^{(l)}\frac{\var^p (\rSigFrq_k - \hat{p}-\est{\chFrq}_k\sym^{(l)})}{\bckPwr + \var^{p}+\abs{\sym^{(l)}}^2\var_k^{\chFrq}}\label{eq:JNEDDataChE}
\end{align}
where $\posSymBlf_k^{(l)}$ is the posterior symbol probability from \eqref{eq:postSym}
but now with $\pdf{\rSigFrq_k|\sym_k;\est{p},\var^p}=\N{\rSigFrq_k;\hat{p}+\est{\chFrq}_k\sym_k,\var^{p}+\abs{\sym_k}^2\var^{\chFrq}_k+\bckPwr}$.
Similarly, the \textit{law of total variance} implies 
\begin{align}
\lefteqn{ \varOp{\impNFrq_k|\rSigFrq_k;\hat{p},\var^{p}} }\nonumber\\
&= \E[\sym_k|\rSigFrq_k]{\varOp{\impNFrq_k|\sym_k,\rSigFrq_k;\hat{p},\var^{p}}} 
+ \varOp[\sym_k|\rSigFrq_k]{\E{\impNFrq_k|\sym_k,\rSigFrq_k;\hat{p},\var^{p}}} \nonumber\\
&= \sum\limits_{l=1}^{\card{\constl}}\posSymBlf_k^{(l)} 
\Bigg[\frac{\var^{p}\left(\bckPwr+\abs{\sym^{(l)}}^2\var_k^{\chFrq}\right)}{\bckPwr + \var^{p}+\abs{\sym^{(l)}}^2\var_k^{\chFrq}}  
+ \big|\E{\impNFrq_k|\sym_k,\rSigFrq_k;\hat{p},\var^{p}}\big|^2
\Bigg] 
\nonumber\\&\qquad  
- \big|\E{\impNFrq_k|\rSigFrq_k;\hat{p},\var^{p}} \big|^2 .
\end{align}

The derivation for pilot tones $k\in \pilotSet$ reduces to the above under $\posSymBlf_k^{(1)}=1$, $\posSymBlf_k^{(l\neq 1)}=0$, and $\sym^{(1)}=\pilot$.
Meanwhile, the derivation for null tones $k\in\nullSet$ is the special case of pilots with $\pilot=0$.

%
%
%
%
%

\ifCLASSOPTIONcaptionsoff
  \newpage
\fi



\bibliographystyle{IEEEtran}
\bibliography{IEEEabrv,refs}
\end{document}